\documentclass[letterpaper,12pt]{article}

\usepackage{tabularx} 
\usepackage{jheppub}
\usepackage{amsmath}  
\usepackage{graphicx} 
\usepackage{subcaption}
\usepackage[utf8]{inputenc}
\usepackage[T1]{fontenc}
\usepackage{lmodern}
\usepackage{tikz}

\usepackage{hyperref}
\def\be{\begin{equation}}
\def\ee{\end{equation}}

\begin{document}

\title{Brick Wall in AdS-Schwarzschild Black Hole: Normal Modes and Emerging Thermality}
\author[a]{Suman Das,}
\author[b]{Somnath Porey,}
\author[b]{Baishali Roy}
\affiliation[a]{Theory Division, Saha Institute of Nuclear Physics, A CI of Homi Bhabha National Institute, 1/AF, Bidhannagar, Kolkata 700064, India.}
\affiliation[b]{Ramakrishna Mission Vivekananda Educational and Research Institute, Belur Math, Howrah-711202, West Bengal, India}
\emailAdd{suman.das[at]saha.ac.in, somnathhimu00[at]gm.rkmvu.ac.in, baishali.roy025[at]gm.rkmvu.ac.in}
\date{\today}


\abstract{
This paper investigates the normal modes of a probe scalar field in a five-dimensional AdS-Schwarzschild black hole with the brick wall boundary condition near the horizon. We employ various techniques
to compute the spectrum and analyze its properties. Our results reveal a linear dependence of the spectrum on the principal quantum number while demonstrating a non-trivial dependence on the angular momentum quantum number. 
We compute the Spectral Form Factor (SFF) and find a dip-ramp-plateau structure, with the slope of the ramp approaching unity as the brick wall nears the horizon. This feature appears to vanish when the horizon size becomes much larger compared to the AdS length. We also observe that as the brick wall approaches the horizon, the poles of the retarded Green's function condense on the real line, leading to an emergent thermal behavior in the boundary theory. This work extends previous studies on lower-dimensional black holes to higher dimensions, providing insights into the connection between black hole microstate models and boundary chaos. Our findings contribute to the ongoing discussions on the information paradox and the nature of black hole interiors in the context of AdS/CFT correspondence.} 

\maketitle





\section{Introduction}

Strongly interacting theories are generally chaotic. While chaos in classical mechanics is well-defined, its counterpart in the quantum world is less clear. Various attempts have been made to bridge these two realms at the semiclassical level \cite{gutzwiller1991chaos, doe1980phase, Berry, MVBerry_1977}, but a comprehensive understanding remains elusive. Over the years, several measures have been developed to quantify quantum chaos, yet ambiguity persists. One notable measure is the Out-of-Time-Ordered Correlator (OTOC) \cite{1969JETP...28.1200L, Maldacena:2015waa}, which measures early-time chaos, whereas the Spectral Form Factor (SFF) \cite{Br_zin_1997, Cotler:2016fpe} measures chaos at late times. Level spacing distribution (LSD) has also been a valuable tool for understanding quantum chaos since the pioneering work of Wigner and Dyson \cite{Wigner_1951, dyson1962statistical}. Although strongly interacting systems are intriguing, they are notoriously difficult to solve. This is where Random Matrix Theory (RMT) universality provides critical insights into the statistical properties of these systems. RMT is universal in the sense that the symmetry class of the Hamiltonian determines the ensemble of the random matrix theory, offering a powerful framework for analyzing complex quantum systems.

Thermality is intrinsically linked with chaos, and it is both intriguing and challenging to understand how thermality arises in a closed, isolated system. Though there are several ideas to support this, including typicality and the concept of a bath (see \cite{Deutsch_2018} for details), the satisfactory mechanism is not yet well understood. One such mechanism is provided by the Eigenstate Thermalization Hypothesis (ETH) \cite{Srednicki_1999, Deutsch_2018}. Furthermore, for a pure state with very large entropy (e.g., large $N$), quantum statistical mechanics shows that the variance of any local correlation function is suppressed by a factor of $e^{-S}$ \cite{Balasubramanian:2007qv}, where S can be identified as the entropy. Therefore, in a strongly interacting system with a large number of degrees of freedom, it is possible that the low-point correlation function can be well approximated by its thermal expectation value.

At first glance, the concepts discussed above may seem disconnected, but they are closely related to black holes and quantum gravity. The AdS/CFT correspondence \cite{Maldacena:1997re} offers a framework that links an interacting conformal field theory (CFT) with quantum gravity in anti-de Sitter (AdS) spacetime. Since its inception, AdS/CFT has provided numerous insights that have deepened our understanding of black holes and their semiclassical properties \cite{Maldacena:2001kr, Barbon:2003aq, Sekino:2008he, VanRaamsdonk:2010pw, Shenker:2013pqa, Maldacena:2013xja, Susskind:2014rva, Almheiri:2014lwa}. However, the questions of smoothness of the horizon and the existence of an interior remain unresolved \cite{Mathur:2009hf, Almheiri:2012rt}. This smoothness is at the heart of the information paradox \cite{Hawking:1975vcx, Hawking:1976ra, Mathur:2009hf}. Despite notable recent progress \cite{Penington:2019npb, Almheiri:2019psf}, the paradox remains unresolved, especially in the Lorentzian picture and for the higher-dimensional black holes.

On the other hand, string theory provides alternatives such as fuzzballs which can bypass this. According to the fuzzball proposal, the horizon is not a smooth place and must be replaced by a complex stringy structure \cite{Lunin:2001fv, Mathur:2005zp, Bena:2007kg, Mathur:2014nja}. For some particular cases, there exists a whole moduli space of such solutions which, when quantized and counted, give rise to perfect agreement with the Bekenstein-Hawking entropy of the black hole \cite{Rychkov:2005ji, Krishnan:2015vha}. Another interesting fact is that, in the low-energy supergravity limit, many of these solutions can be written as explicit metrics called microstate geometries (see \cite{Bena:2022rna} for a review). These are perfectly regular geometries and behave similarly to black holes from a distance, but cap off smoothly near the horizon. Although there are many criticisms (e.g., \cite{Raju:2018xue}) regarding the geometric picture, the main takeaway (for our purpose) is that string theory (as a UV-complete theory) can provide mechanisms that support a structure near the horizon\footnote{Notably, long ago, 't Hooft considered an ad-hoc model \cite{tHooft:1984kcu} by placing a brick wall at a Planck distance away from the horizon and quantizing scalar fields in this geometry, which reproduced the entropy as a one-loop effect (see also \cite{Burman:2023kko} for a microcanonical version of the same).}.

Black holes are expected to be chaotic objects as their boundary duals are strongly coupled interacting theories. This can be seen by computing the out-of-time-order correlators (OTOCs) in black hole geometries \cite{Shenker:2013pqa, Maldacena:2015waa}. Moreover, it has been shown that black holes are maximally chaotic in the sense that the Lyapunov exponent saturates the chaos bound,
\begin{equation}
\lambda= \frac{2\pi}{\beta},
\end{equation}
where $\beta$ is the inverse Hawking temperature of the black hole. It is important to remember that OTOCs measure early-time chaos, and there is evidence (e.g., \cite{Rozenbaum:2019nwn}) that early-time chaos does not always imply chaos for all times. This raises the question: How can we demonstrate that the system remains chaotic at late times? Tools such as the SFF and LSD require knowledge of the spectrum, necessitating the quantization of the black hole system—a challenging problem \footnote{In two-dimensional JT gravity, which is dual to the SYK model, it has been shown that the SFF exhibits a linear ramp with a slope of one, and the LSD follows the Wigner-Dyson distribution \cite{Cotler:2016fpe}.}. A simpler and more tractable question might be: Can we detect any hint of late-time chaos in the probe sector?

A naive probe calculation results in complex-valued quasi-normal modes, leading to a vanishing SFF at late times, a manifestation of the information paradox as discussed in \cite{Maldacena:2001kr}. To circumvent this problem, \cite{Das:2022evy, Das:2023ulz} proposed placing a brick wall in the geometry \footnote{This is motivated by the fuzzball picture. Note that most actual fuzzball geometries are for extremal black holes, making the notion of chaos less meaningful there (see \cite{chen2024bpschaos} for recent progress).} and quantizing a probe scalar field to obtain normal modes as a function of the wall's position. It was shown that when the brick wall is placed very near to the horizon, the SFF constructed from these normal modes exhibits a clear Dip-Ramp-Plateau structure with a linear ramp of slope one in the log-log plot. Although the LSD is not of the Wigner-Dyson type (despite exhibiting level repulsion), a WD-type LSD was achieved by imposing fluctuations in the stretched horizon, which may be natural in constructing a more realistic toy version of actual fuzzballs \cite{Das:2023ulz}. This approach was generalized for rotating BTZ geometry in \cite{Das:2023xjr}. It has been argued in \cite{Das:2022evy, Das:2023ulz} that this behavior is a generic feature of any non-extremal black hole, with supporting calculations in the Rindler $\times S^1$ framework. However, no explicit calculations have been performed for higher-dimensional black holes. This article fills this gap by computing the normal modes of a probe scalar field in a five-dimensional AdS Schwarzschild black hole.

This study is intriguing for several reasons. Firstly, in three dimensions, gravity is non-dynamical as the degrees of freedom of the graviton are zero, whereas in higher dimensions, gravity is fully dynamical. Secondly, the AdS/CFT correspondence is much better understood for AdS$_5$-CFT$_4$. For example, the CFT is a $3+1$ dimensional $\mathcal{N}=4$ super Yang-Mills (SYM) theory, where the glueball operators Tr$(F_{\mu \nu} F^{\mu\nu})$ and Tr$(F_{\mu \nu} \tilde{F}^{\mu\nu})$ are dual to the dilaton and axions, which are minimally coupled scalar fields in the bulk AdS. Thus, we can relate the bulk results to a more realistic four-dimensional strongly interacting theory.

The organization of the paper is as follows: In the next section (Section [\ref{sec:2}]), we describe the setup and identify the radial equation as the Heun equation. Subsequently, we attempt to solve the Heun equation using different methods and extract the normal modes. In Section [\ref{sec:3}], we leverage the fact that the BPZ equation in Liouville CFTs also satisfies the Heun equation. By using crossing symmetries, one can relate the solutions around different singular points. Using the solutions around different singular points, we first compute the quasinormal modes with the ingoing boundary condition in Section [\ref{sec:3.1}], and then extract the normal modes in Section [\ref{sec:3.2}]. In Section [\ref{sec:4}], we use the WKB approximation method to find the normal modes with Dirichlet boundary conditions on the probe field near the horizon. Additionally, we demonstrate that the SFF in this case exhibits a dip-ramp-plateau behavior, with the slope of the ramp approaching one as the Dirichlet wall moves closer to the horizon. In Section [\ref{sec:5}], we use another perturbative method to solve the Heun equation and compute the normal modes. In Section [\ref{sec:6}], we show the emergence of a branch cut-like structure in the Green's function of the probe field as the wall is moved closer to the horizon. Finally, in Section [\ref{sec:7}], we conclude with a discussion and outline a few possible future directions. Additionally, we have provided three appendices ([\ref{apnAA}], [\ref{apnA}], [\ref{apnB}]) to make the paper self-contained.


\section{The Setup: Scalar Field in AdS Schwarzschild Black hole} \label{sec:2}

Let's consider a probe scalar field of mass $\mu$ in  the five-dimensional AdS-Schwarzschild black hole geometry,
\begin{equation}
    ds^2=-f(r) dt^2+\frac{dr^2}{f(r)}+r^2 \, d\Omega_3^2 \, ,
\end{equation}
with $f(r)= \left(1-\frac{r_H^2}{r^2}\right)(r^2+r_H^2+1)$, where $r_H$ represents the position of the horizon (we have set the AdS length $l$ to $1$). The mass of the black hole, M is related to the horizon as,
\begin{equation}
    r_H^2=\frac{1}{2} \left( \sqrt{1+\frac{32 G_N M}{3 \pi}} -1 \right).
\end{equation}
As mentioned in the introduction, our `measuring probe' is a  scalar field $\Phi$ in this black hole background which satisfies the following Klein-Gordon equation:
\begin{equation}\label{eom1}
    \Box \Phi\equiv \frac{1}{\sqrt{|g|}}\partial_{\nu}\left(\sqrt{|g|}\partial^{\nu}\Phi\right)=\mu^2 \Phi.
\end{equation}
Where $\mu$ is the mass associated with the scalar field. The conformal dimension of the boundary operator dual to $\Phi$ is given by,
\begin{eqnarray}
    \mu=\sqrt{\Delta(\Delta-4)} \, .
\end{eqnarray}
Since the metric is time-independent and spherically symmetric, we can use the ansatz
\begin{equation}
    \Phi(t,r,\Omega) \sim e^{-i\omega t} Y_{l,\vec{m}}(\Omega)\phi_{\omega l}(r)\, ,
\end{equation}
where $Y_{l,\vec{m}}(\Omega)$ represent spherical harmonics on $S^3$. These harmonics satisfy the following equation:
\begin{equation}
    \nabla^2_{S^3} Y_{l,\vec{m}} =-l(l+2) Y_{l,\vec{m}},
\end{equation}
where $\nabla^2_{S^3}$ is the Laplacian on $S^3$ and $\vec{m}=(m, m')$ can go from $-l/2$ to $l/2$ (see \cite{Festuccia:2005pi}). With this, \eqref{eom1} can be written as (to simplify our notation, we will now write $\phi_{\omega l}(r)$ as $\phi(r)$),
\begin{equation}\label{eom2}
    \frac{1}{r^3} \frac{d}{dr} \left( r^3 f(r) \frac{d\phi(r)}{dr} \right)+ \left( \frac{\omega^2}{f(r)}-\frac{l(l+2)}{r^2}-\mu^2  \right) \phi(r)=0.
\end{equation}
This is a Heun's equation which can be written in the well-known normal form by using the following transformations,
\begin{align}
    z &=\frac{r^2}{r_H^2+r^2+1}\, ,\\
    \phi(r) &= z^{-\frac{1}{2}} (z-\tilde{t})^{-\frac{1}{2}} (1-z)^{\frac{1}{2}} \chi(z)\, .
\end{align}
In this new $z$ coordinate, the horizon is located at $z=\tilde{t}=\frac{r_H^2}{2r_H^2+1}$ and the boundary is at $z=1.$ The radial equation is,
\begin{equation}\label{eom3}
    \left( \partial_z^2+\frac{\frac{1}{4}-a_1^2}{(z-1)^2}- \frac{\frac{1}{2}-a_0^2-a_1^2-a_{\tilde{t}}^2+a_{\infty}^2+u}{z(z-1)}+ \frac{\frac{1}{4}-a_{\tilde{t}}^2}{(z-\tilde{t})^2}+\frac{u}{z(z-\tilde{t})}+ \frac{\frac{1}{4}-a_0^2}{z^2} \right)\chi(z)=0.
\end{equation}
The various parameters appearing in equation \eqref{eom3} are detailed in Appendix \ref{apnB}.

Near horizon behaviour of $\chi(z)$ is the following
\begin{align}\label{nhor}
    \chi_{\text{hor}}(z) &= c_1 \, (\tilde{t}-z)^{\frac{1}{2}+a_{\tilde{t}}}+c_2 \, (\tilde{t}-z)^{\frac{1}{2}-a_{\tilde{t}}}+ \ldots \nonumber \\
    &= c_1 \chi^{(\tilde{t})}_{+}(z)+ c_2 \chi^{(\tilde{t})}_{-}(z) +\ldots
\end{align}
Here, the first term corresponds to the outgoing mode and the second term to the ingoing mode near the horizon. The constants $c_1$ and $c_2$ are arbitrary.\\
Similarly, Near boundary behavior can be expressed as,
\begin{align}\label{nbdry}
    \chi_{\text{bdry}}(z) &= c_3\left( \frac{1-z}{1+r_H^2}  \right)^{\frac{1}{2}+a_1}+ c_4 \, \left( \frac{1-z}{1+r_H^2}  \right)^{\frac{1}{2}-a_1}  +\ldots \nonumber \\
    &= c_3 \, \chi^{(1)}_{+}(z)+ c_4 \, \chi^{(1)}_{-}(z) +\ldots
\end{align}
Where the first term corresponds to the normalizable mode and the second term is non-normalizable.

Since the radial equation is a Heun's equation, a closed-form solution in terms of known functions is not available. However, we can still use techniques from Liouville CFT to write the connection formulas that relate the solutions around various singular points. These connection formulas will aid us in finding the normal modes, as we will discuss in the next section.

\section{Using CFT techniques}\label{sec:3}

In Liouville CFT, the BPZ equation reduces to the Heun's equation in normal form in the semi-classical limit. The solutions of this equation give conformal blocks associated with different channels. Therefore, using the crossing symmetries of the conformal blocks, we can determine the relations between solutions around different singular points \cite{Bonelli:2022ten}. 
Let's focus first on the computation of quasi-normal modes and Green's function in this black hole metric.

\subsection{Ingoing boundary condition and quasinormal modes} \label{sec:3.1}

The natural boundary condition near the horizon of a black hole is ingoing boundary condition which implies $c_1=0$. So,
\begin{equation}\label{lcft2}
     \chi_{\text{hor}}(z) = c_2 \,  \chi_{-}^{(\tilde{t})} (z) .
\end{equation}

Using the connection formulas, we can write the near-horizon solution in terms of the solution near the boundary. Let's write the relations between the near-horizon and near-boundary solutions in the following way \footnote{The exact expressions of $a_{11}, a_{22}, b_{11}, b_{22}$ are given in the appendix \ref{apnB}.}:

\begin{eqnarray}\label{lcft5}
    \chi^{(\tilde{t})}_{+} = a_{11} \chi^{(1)}_{+} + a_{22} \chi^{(1)}_{-} \, , \\
    \chi^{(\tilde{t})}_{-} = b_{11} \chi^{(1)}_{+} + b_{22} \chi^{(1)}_{-} \, .
\end{eqnarray}
Using \eqref{lcft5} in \eqref{lcft2}, we get


\begin{equation}
    \chi_{\text{hor}} (z)= c_2 \left( b_{11} \chi^{(1)}_{+}+ b_{22} \chi^{(1)}_{-}   \right)\, .
\end{equation}
To find the quasi-normal modes we need to impose normalizablity at the boundary, which implies,
\begin{eqnarray}
    b_{22}=M_{-+}(a_{\tilde{t}}, a; a_0) M_{--}(a, a_1; a_{\infty})+M_{--}(a_{\tilde{t}}, a; a_0) M_{+-}(a, a_1; a_{\infty})  \tilde{t}^{-2a} e^{\partial_a F}=0
\end{eqnarray}
This can be solved to extract the quasinormal modes of the black hole (see \cite{Dodelson:2022yvn, Jia:2024zes} for the details). Following the Son-Starinets prescription \cite{Son:2002sd}, the retarded Green's function can be expressed as the ratio of the normalizable to non-normalizable modes as follows,
\begin{equation} \label{greenfn}
    G_{R}^{BH}(\omega, \lambda)= \frac{b_{11}}{b_{22}}.
\end{equation}
%

\subsection{Dirichlet boundary condition and normal modes}\label{sec:3.2}
\begin{figure}
\begin{subfigure}{0.47\textwidth}
    \centering
    \includegraphics[width=\textwidth]{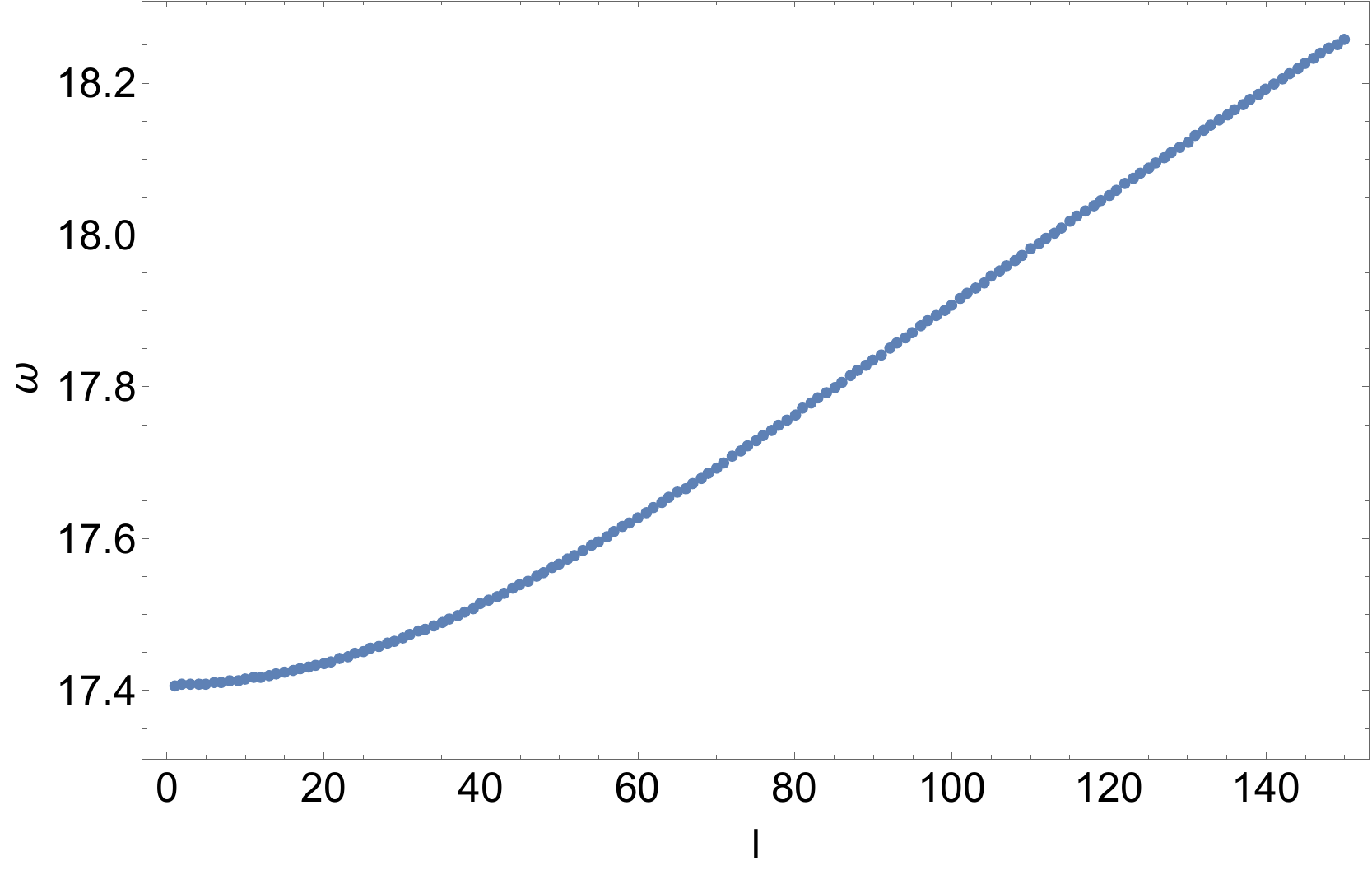}
    \end{subfigure}
    \hfill
    \begin{subfigure}{0.47\textwidth}
    \includegraphics[width=\textwidth]{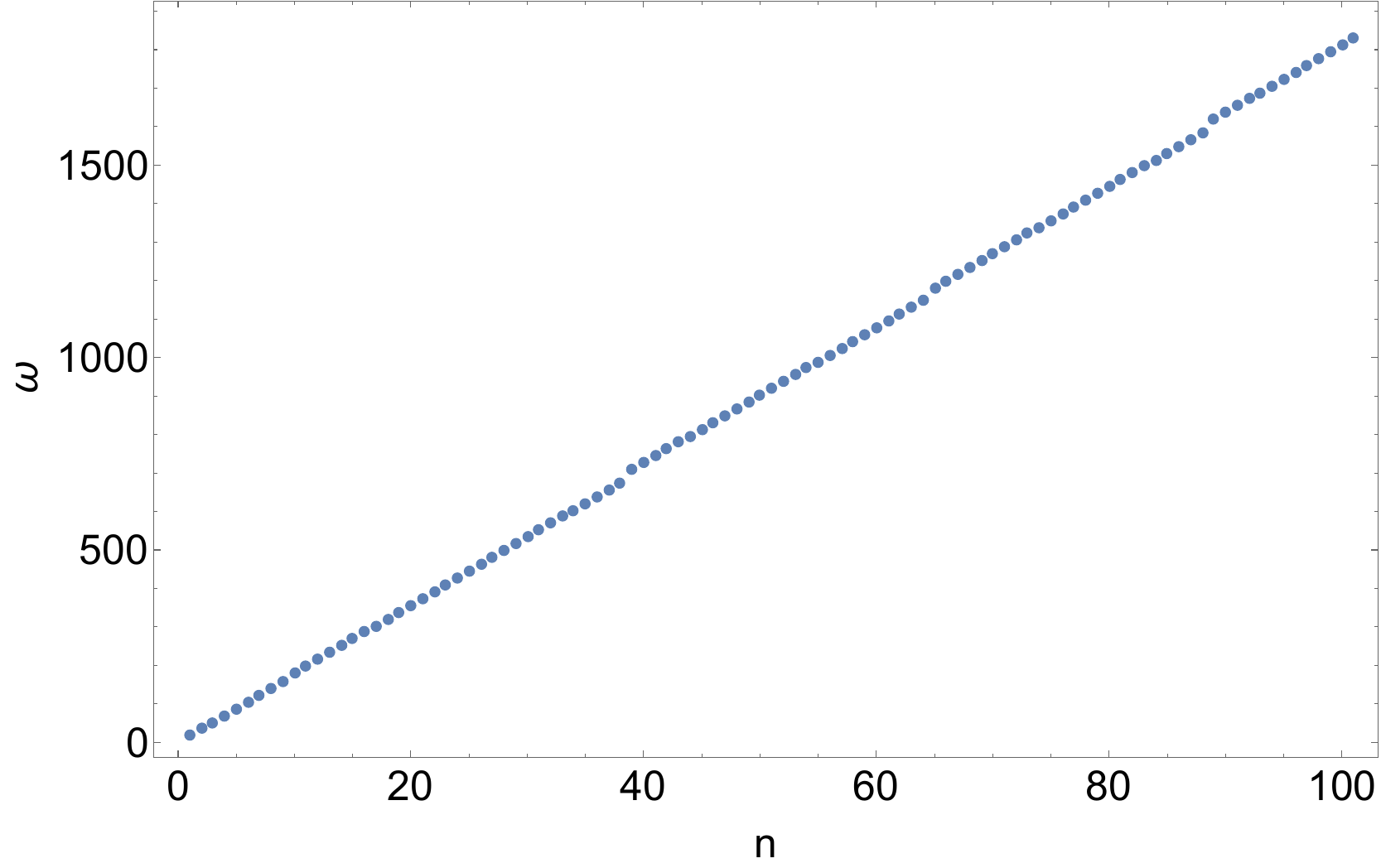}
    \end{subfigure}
    \caption{Spectrum along $l$ (left) and $n$ (right) directions obtained by solving \eqref{quantLcft}. The parameters used are $r_H=100, \Delta=4.1$ and $\epsilon_0=10^{-30}$. Although the spectrum has a trivial linear dependence on $n$-quantum number, along $l$ direction it is non-trivial.}
    \label{spectrum_LCFT}
\end{figure}

\begin{figure}
\begin{subfigure}{0.47\textwidth}
    \centering
    \includegraphics[width=\textwidth]{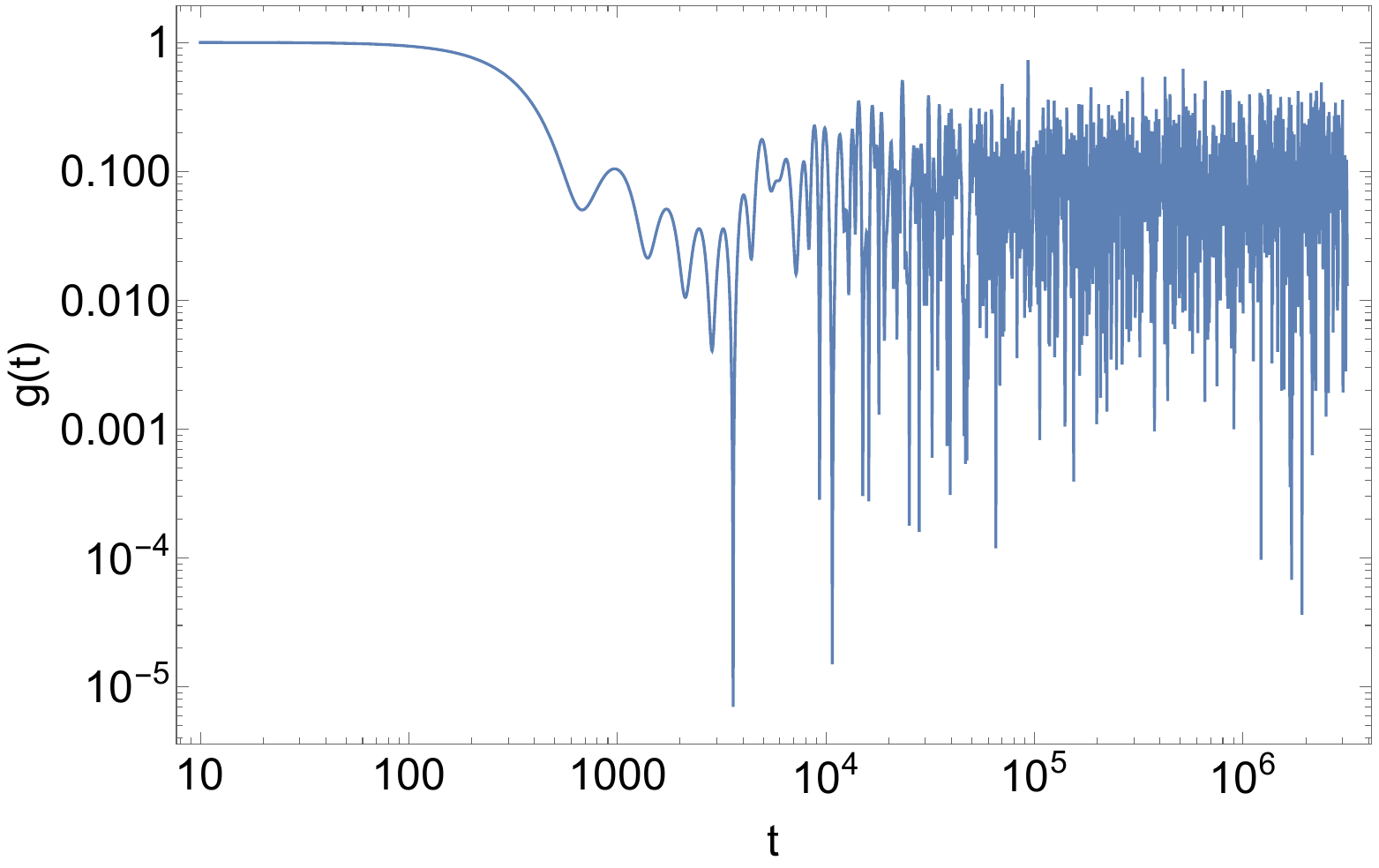}
    \end{subfigure}
    \hfill
    \begin{subfigure}{0.47\textwidth}
    \includegraphics[width=\textwidth]{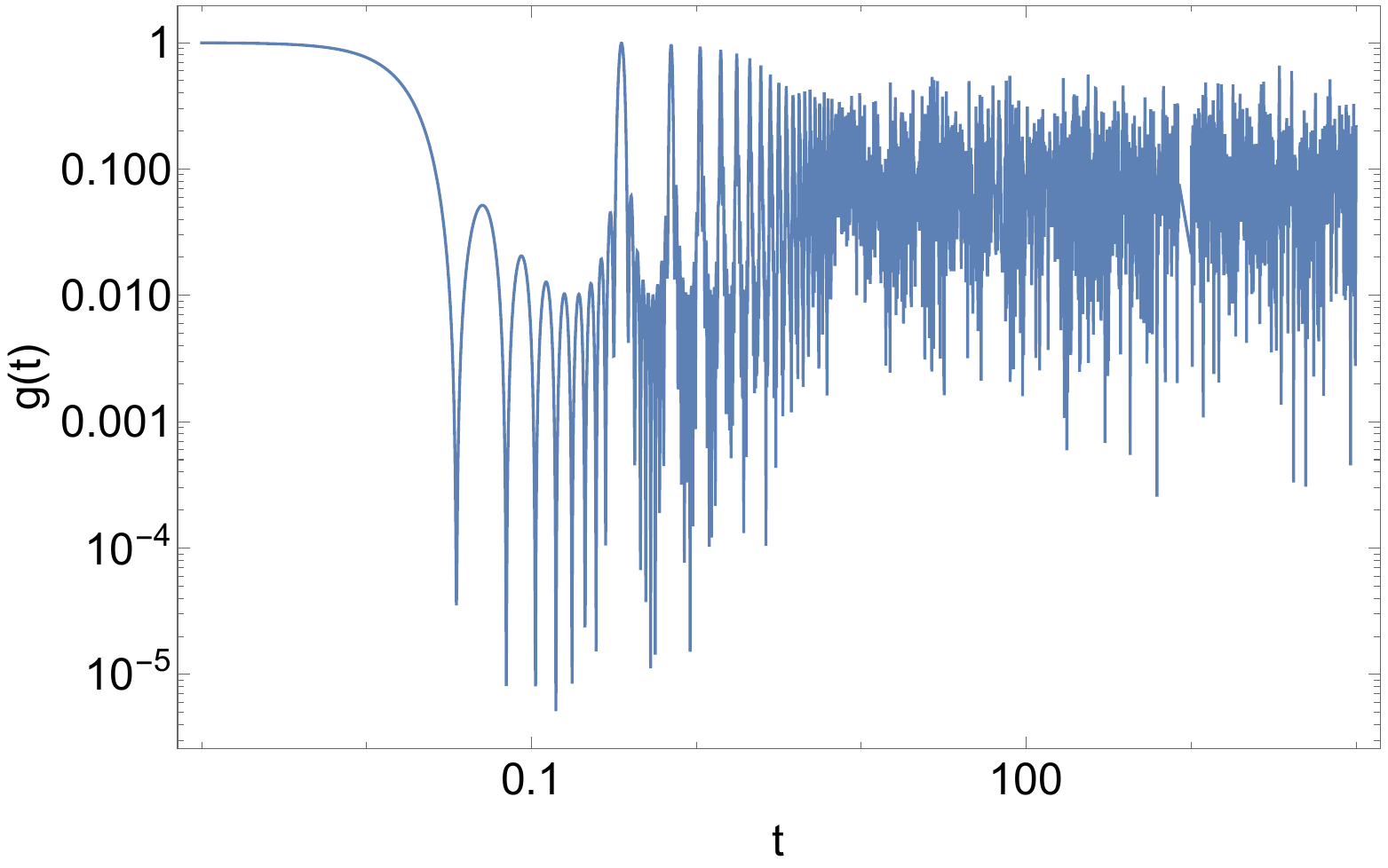}
    \end{subfigure}
    \caption{This figure illustrates the SFF of the modes presented in Figure \ref{spectrum_LCFT}, along the $l$-direction (left) and $n$-direction (right), with $\beta$ fixed to zero.}
    \label{sff_ln}
\end{figure}

\begin{figure}
    \centering
    \includegraphics[width=.55\textwidth]{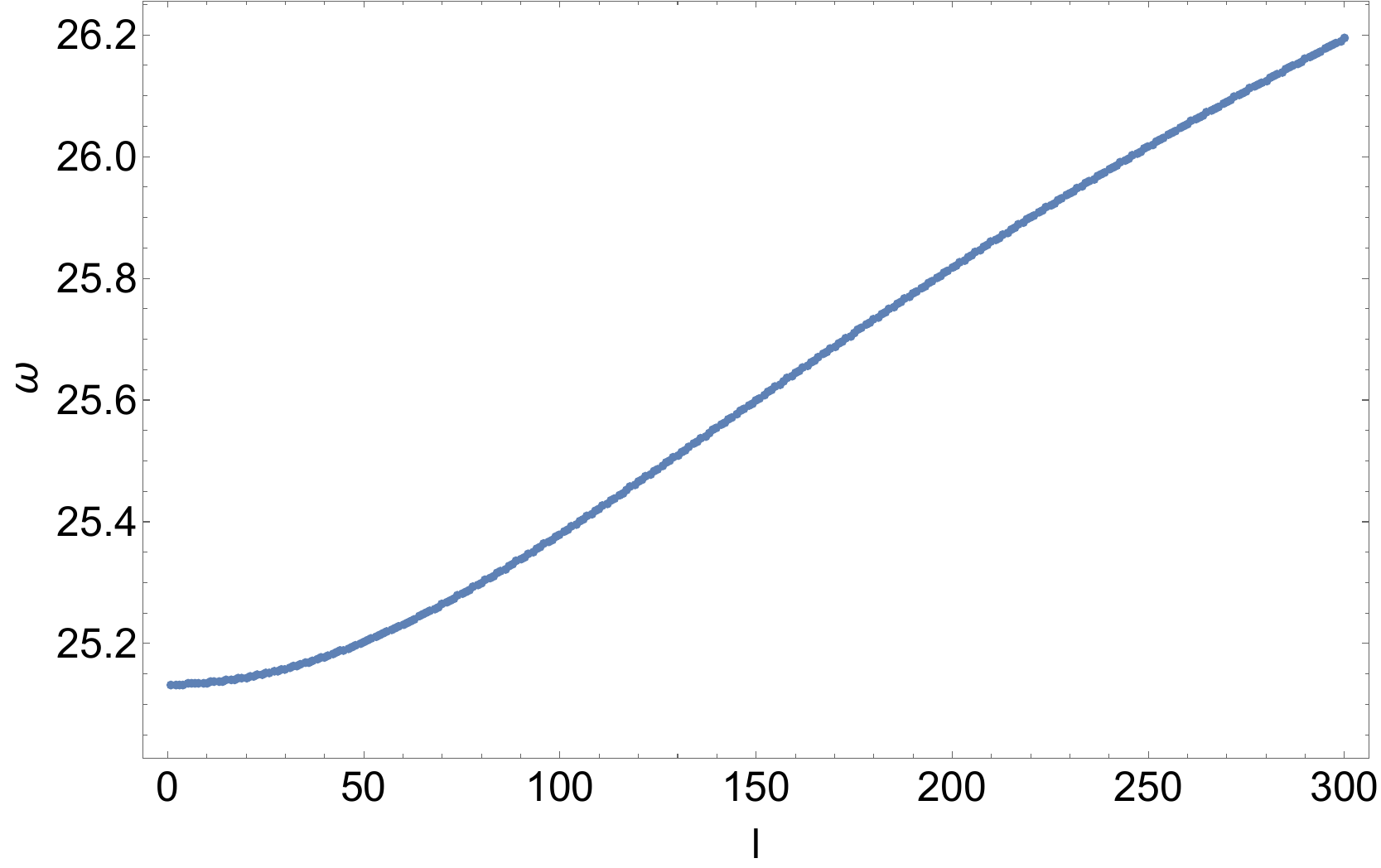}
   \caption{Normal modes along $l$ direction for BTZ black hole for $r_H/l=100 \gg1$.}
    \label{BTZ_m}
\end{figure}

Instead of imposing an ingoing boundary condition near the horizon, we now consider the Dirichlet boundary condition $\Phi(z_0) = 0$ at some $z_0 > \tilde{t}$. This boundary condition is motivated by the desire to connect the brick wall-type model of 't Hooft with the fuzzball (microstate geometry) picture. It is important to note that, in principle, the brick wall can be placed at any location since it is an ad hoc boundary condition within the realm of classical General Relativity. However, as noted in works by 't Hooft \cite{tHooft:1984kcu} and \cite{Das:2022evy, Das:2023ulz, Banerjee:2024dpl, Krishnan:2023jqn, Burman:2023kko}, interesting physics emerges when the brick wall is very close to the horizon. Therefore, we take $z_0 \rightarrow \tilde{t}$. In this region Dirichlet boundary condition implies,
\begin{equation}
     \chi_{\text{hor}}(z_0) =c_1 \,  \chi_{+}^{(\tilde{t})} (z_0) + c_2 \,  \chi_{-}^{(\tilde{t})} (z_0) = 0 \, ,
\end{equation}
which implies,
\begin{align}\label{r12}
    R_{c_1c_2} = \frac{c_1}{c_2} &= -\frac{\chi_{-}^{(\tilde{t})} (z_0)}{\chi_{+}^{(\tilde{t})} (z_0)} \nonumber \\
    &= -\frac{(\tilde{t}-z_0)^{\frac{1}{2}-a_{\tilde{t}}}}{(\tilde{t}-z_0)^{\frac{1}{2}+a_{\tilde{t}}}} \nonumber \\
    &= -(\tilde{t}-z_0)^{-2a_{\tilde{t}}}.
\end{align}
Then the near-horizon solution comes out to be,
\begin{equation}\label{diri}
    \chi^{(\tilde{t})}(z)= c_2(R_{c_1c_2} \,  \chi^{(\tilde{t})}_{+}(z)+\chi^{(\tilde{t})}_{-}(z)).
\end{equation}
Using \eqref{lcft5} in \eqref{diri}, we can relate the near-horizon and near-boundary solutions as follows:
\begin{equation}\label{eq00}
    \chi^{(\tilde{t})}= c_2 \left((R_{c_1c_2} \, a_{11}+ b_{11})\chi^{(1)}_{+}+ (R_{c_1c_2} \, a_{22}+ b_{22})\chi^{(1)}_{-}   \right).
\end{equation}
Then normalizable condition at the boundary implies,
\begin{equation}\label{eq11}
    R_{c_1c_2} \, a_{22}+ b_{22}=0.
\end{equation}
Substituting \eqref{r12} in \eqref{eq11},
\begin{align}\label{quan0}
    \frac{b_{22}}{a_{22}} &=(\tilde{t}-z_0)^{-i\omega  \frac{r_H}{1+2r_H^2}} \nonumber\\
     &=\epsilon_0^{-i\omega  \frac{r_H}{1+2r_H^2}}.
\end{align}
Here $\epsilon_0$  measures the distance between the horizon and the brick wall. In Appendix \ref{apnB}, we have shown that the absolute value of  $\frac{b_{22}}{a_{22}}$ is always one, i.e., the absolute part is trivially satisfied \footnote{\eqref{quan0} can be written as $\frac{b_{22}}{a_{22}} \epsilon_0^{i\omega  \frac{r_H}{1+2r_H^2}}=1$, where the left-hand side is a pure phase and can be labeled as $e^{-i \theta(\omega, l)}=1$ .}. Therefore, what we have is the argument part, which is
\begin{equation}
    \text{Arg} \left(\frac{b_{22}}{a_{22}} \right) =\text{Arg} \left(\epsilon_0^{-i\omega  \frac{r_H}{1+2r_H^2}} \right)+ \text{Arg} (1),
\end{equation}
which can be simplified as,
\begin{equation}\label{quantLcft}
    \text{Arg} \left( \frac{b_{22}}{a_{22}} \right)+ \frac{\omega \, r_H}{(1+2r_H^2)} \log \epsilon_0 = 2 n \pi,  \hspace{2cm} \text{where} \, n\in \mathbf{Z}.
\end{equation}
The solution of this equation provides the required normal modes, which are real valued and labeled by two quantum numbers, $n$ and $l$. We have solved this equation using \textit{Mathematica}. The behaviour of the normal modes along the $l$ and $n$ directions is presented in Figure \ref{spectrum_LCFT}. Although the spectrum exhibits a non-trivial dependence on the $l$-quantum number, along the $n$ direction it is almost linear, as observed for the BTZ black hole in \cite{Das:2022evy}. The figure clearly shows that the dependence of the modes on the $l$-direction is much slower compared to the $n$-direction. This quasi-degeneracy along $l$-directions is the key feature that underpins the emergence of interesting non-trivial physics. The corresponding spectral form factor (SFF) [see Appendix \ref{apnAA} for the definition] is presented in Figure \ref{sff_ln}. Unlike the BTZ case, the SFF for the modes along the $l$-direction does not exhibit a linear ramp. This deviation arises from the absence of logarithmic growth in the low-lying modes of the spectrum along $l$. The observed $l$-dependence of the modes raises a natural question: why do these modes differ significantly from those of the BTZ black hole in \cite{Das:2022evy, Das:2023xjr}? This difference is not primarily due to the change in dimensionality (from $3$d to $5$d spacetime) but is instead attributed to the condition $r_H \gg l$. To substantiate this claim, a comparative plot is provided in Figure \ref{BTZ_m}.

\begin{figure}
\begin{subfigure}{0.47\textwidth}
    \centering
    \includegraphics[width=\textwidth]{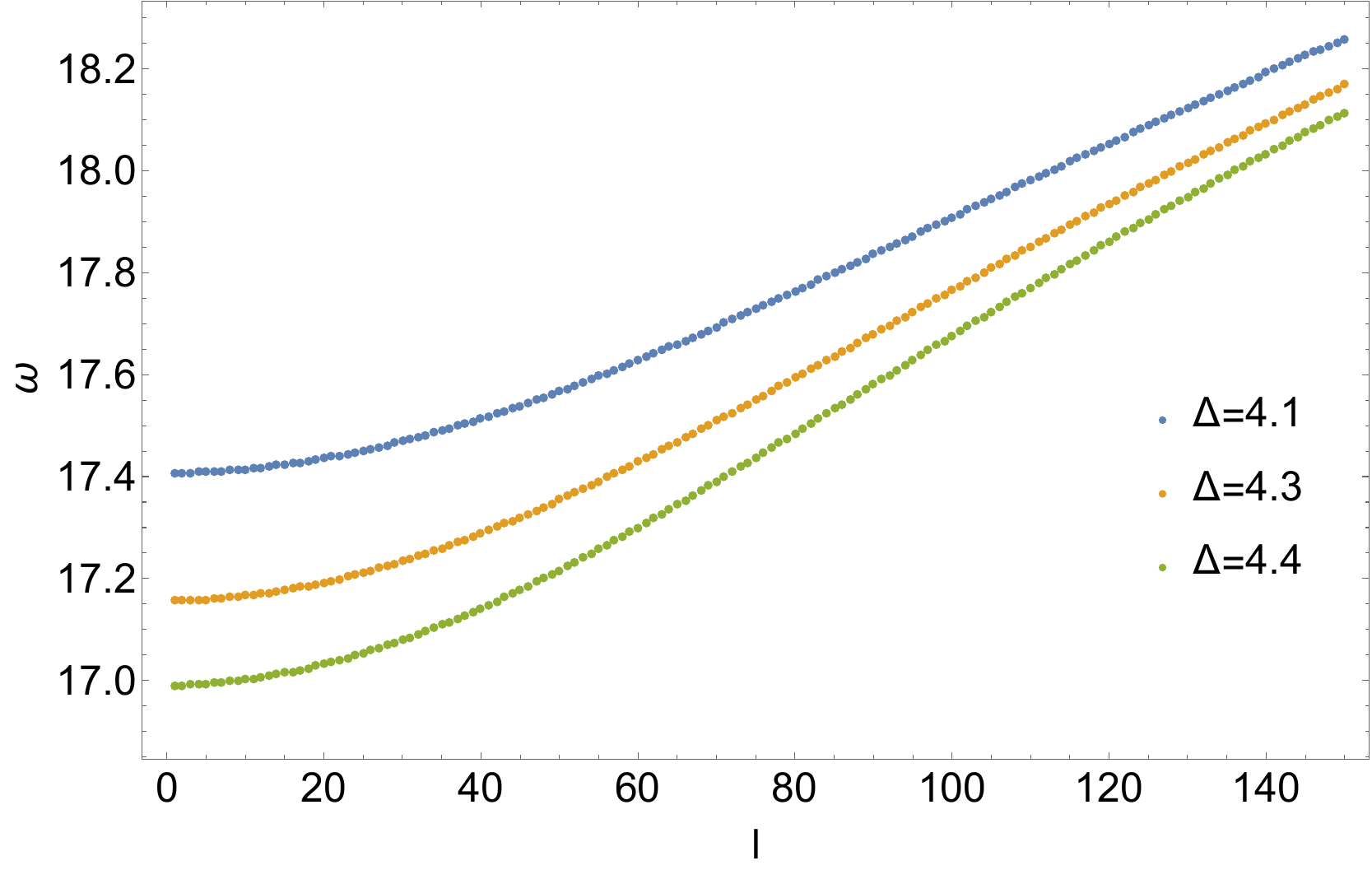}
    \end{subfigure}
    \hfill
    \begin{subfigure}{0.47\textwidth}
    \includegraphics[width=\textwidth]{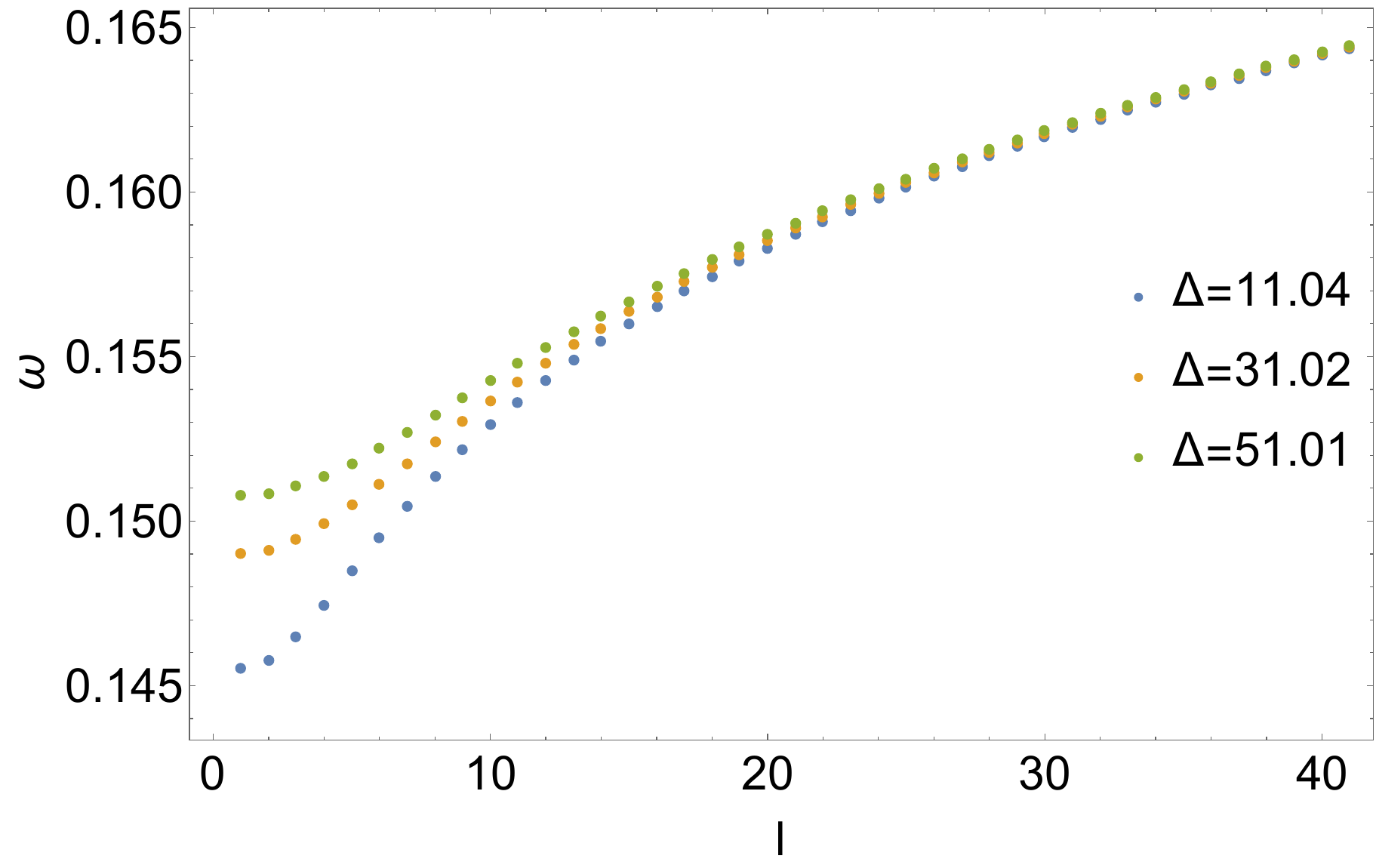}
    \end{subfigure}
    \caption{Left: The $\Delta$ dependence of the normal modes along the $l$ direction is depicted. Right: Similar plots are presented for the BTZ black hole, demonstrating the $\Delta$ dependence of the normal modes along the $m$ direction.}
    \label{delta_dep}
\end{figure}

In Figure \ref{delta_dep} (left), we present the $\Delta$ dependence of the normal modes along $l$-direction. For smaller values of $l$, $\omega$ varies significantly with different $\Delta$. However, as $l$ increases, these differences gradually diminish, and the modes tend to converge onto a single curve. A similar trend is observed for the normal modes of the BTZ black hole, as depicted on the right of Figure \ref{delta_dep}. Notably, there is a key difference between the two cases: in the BTZ black hole, the modes increase with increasing $\Delta>d$, whereas for the AdS-Schwarzschild black hole, the opposite trend is observed, with the modes decreasing as $\Delta$ increases.

Due to the complexity of the various functions appearing in \eqref{quantLcft}, we were unable to solve for larger values of $l$ and $n$ and also for smaller value of $r_H/l$. More advanced numerical techniques are required to extend the solution to larger $l$ and $n$. Nonetheless, we can explore various approximate methods to solve the radial equation, as discussed in the next two sections.
\section{WKB Approximation Method}\label{sec:4}

In this section we will try to solve the radial equation using the WKB method \footnote{For a review of the WKB approximation method, look into appendix [\ref{apnA}].}. For this, we will write \eqref{eom2} in Schrödinger form to get an effective potential, and then we will find the bound states of the potential which will essentially correspond to the normal modes of the scalar perturbation.\\

\noindent In general, an equation of the following form,
\begin{equation}
    a(r) \frac{d^2\phi(r)}{dr^2}+b(r) \frac{d \phi(r)}{dr}+c(r) \phi(r)=0\, ,
\end{equation}
can be written in the form of Schrödinger equation (with $\phi(r)=g(r)\psi(r)$) as,
\begin{equation}
    \frac{d^2\psi(r)}{dr^2}+V(r) \psi(r)=0\, ,
\end{equation}
where,
\begin{equation}
    V(r)=\frac{1}{4 a^2} \big( b^2 - 2 b \, a' + 2 a \left( b'- 2 c \right)  \big).
\end{equation}

\begin{figure}
    \centering
    \includegraphics[width=.55\textwidth]{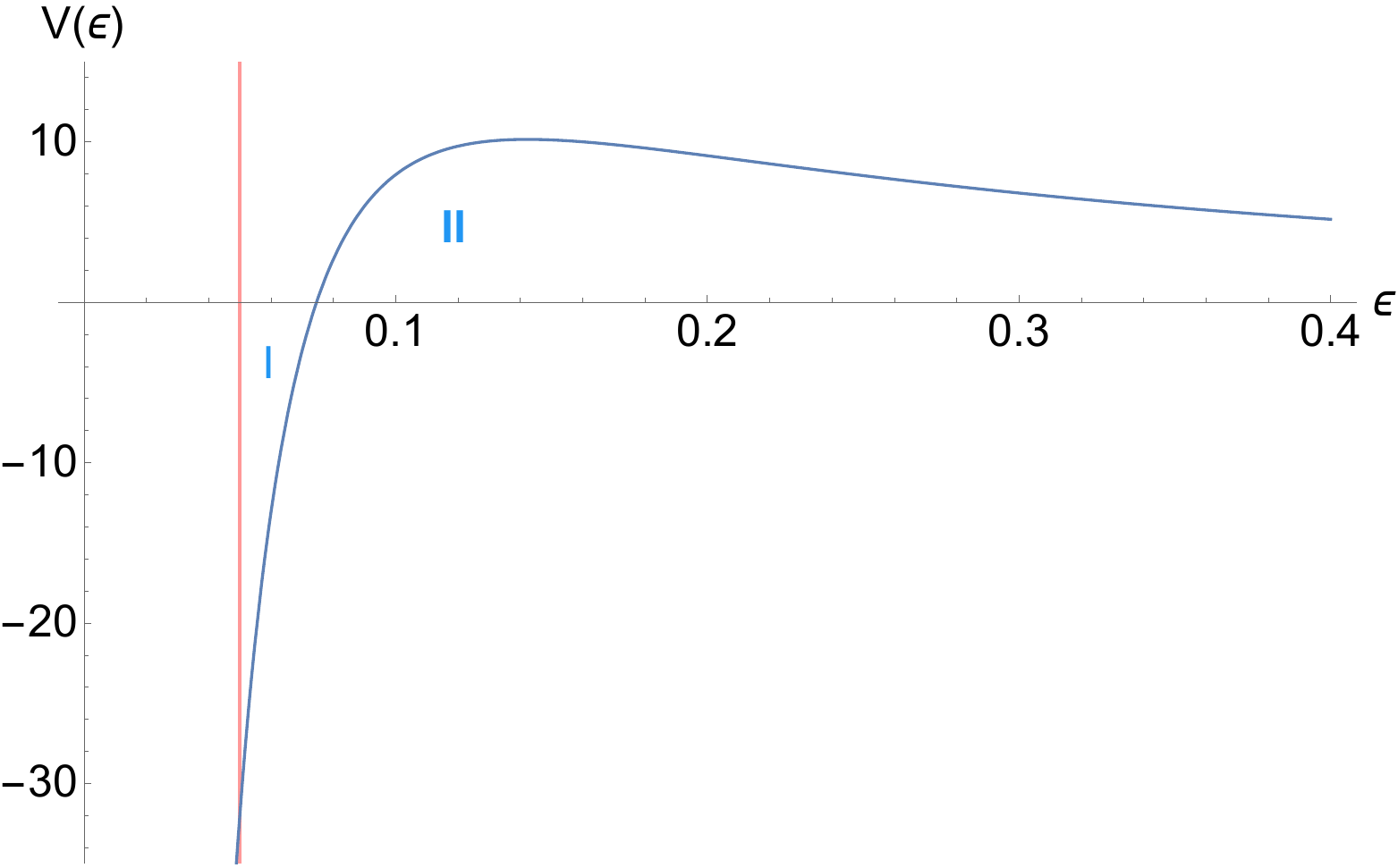}
   \caption{Generic structure of the effective Schrödinger like potential $V(r)$ in \eqref{pot1}. Here $l=3, \omega=0.3$ and $r_H=1$.}
    \label{WKB_pot1}
\end{figure}

\begin{figure}
    \centering
    \includegraphics[width=.55\textwidth]{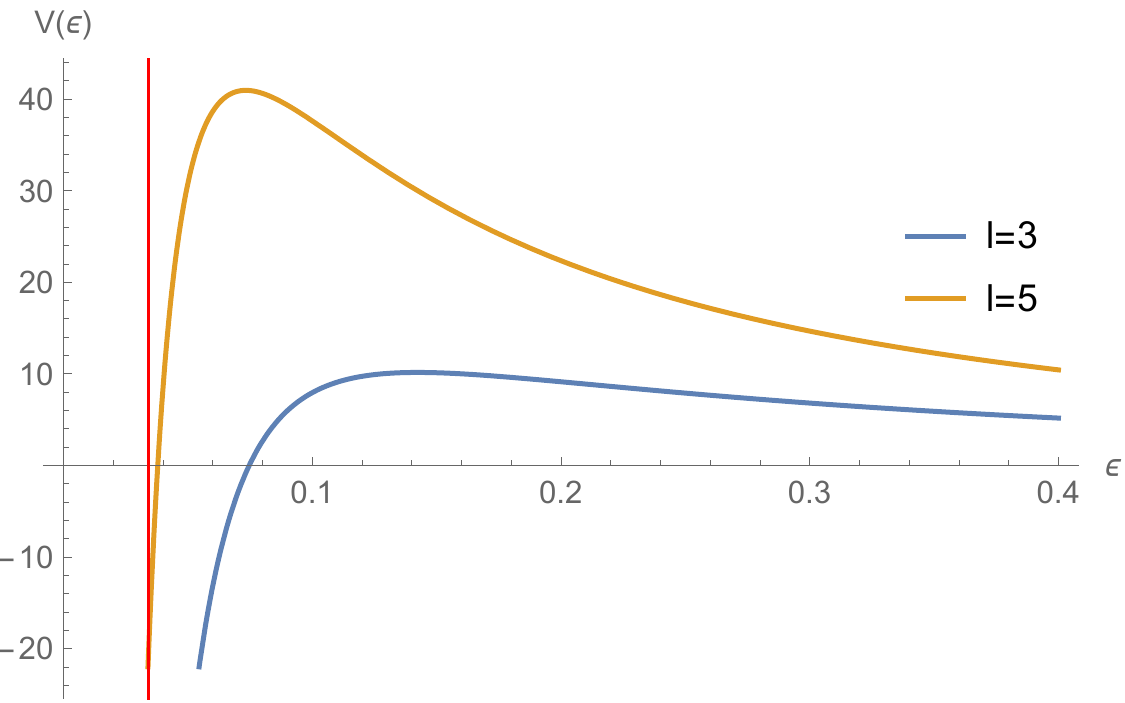}
   \caption{Generic structure of the effective Schrödinger like potential $V(r)$ in \eqref{pot1}. Here $l=3, \omega=0.3$ and $r_H=1$.}
    \label{WKB_pot11}
\end{figure}

Though the general form of the potential is complex and not very enlightening to write down explicitly, some general observations are interesting to note.  For large $r$, $V(r)$ behaves as,
\begin{equation}
    \lim_{r \to \infty} V(r)= \frac{(d^2-1)+4\mu^2}{4r^2}+ O \left(\frac{1}{r^4} \right) ;
\end{equation}
which is always positive. As $r$ approaches to $r_H$, $V(r)$ goes to $-\infty$.  Additionally, there is a turning point at some $r_c>r_H$ where $V(r)$ changes sign. The general structure of $V(r)$ is depicted in Figure \ref{WKB_pot1}. The Dirichlet boundary condition $\phi(r_0)=0=\psi(r_0)$ implies the existence of an infinite potential barrier at the location of the brick wall ($r=r_0$), represented by a red vertical line in Figure \ref{WKB_pot1}. The task is then to find the bound states of this potential well. 
An interesting point to note is that as the angular momentum $l$ increases, the height of the bump after the turning point also increases (see Figure \ref{WKB_pot11}) which is an artifact of the presence of the angular momentum barrier. For a given $r_0$, we can tune $l$  in such a way that no potential well forms, which corresponds to scattering states. These modes are not trapped from the perspective of a boundary observer due to their large angular momentum. We are not interested in those modes and only consider the bound states. \\
The expression for $V(r)$ with $\mu=0$ and $r_H=1$ is given by
\begin{equation}\label{pot1}
    V(r)=\frac{r^6 \left(4 l (l+2)-4 \omega^2+22\right)+(4 l (l+2)-57) r^4-4 (2 l (l+2)+3) r^2+15 r^8-4}{4 r^2 \left(r^4+r^2-2\right)^2}.
\end{equation}
As shown in the appendix (see also \cite{Das:2023xjr},\cite{Bena:2019azk}), to determine the spectrum, we need the closed-form expression of $\int_{r_0}^{r_c} \sqrt{|V(r)|}$, which is not feasible for \eqref{pot1}. However, this does not mark the end of the investigation. Interesting physics occurs when the position of the brick wall is very close to the horizon. We will further assume that the turning point is also very close to the horizon. For the second condition, we need $\frac{\omega}{l}$ not to be very large, meaning we will focus on low-lying modes for a fixed $l$ (i,e, low $n$) which is also physical as long as we are in the probe limit. 

\begin{figure}
     \centering
     \begin{subfigure}[b]{0.45\textwidth}
         \centering
         \includegraphics[width=\textwidth]{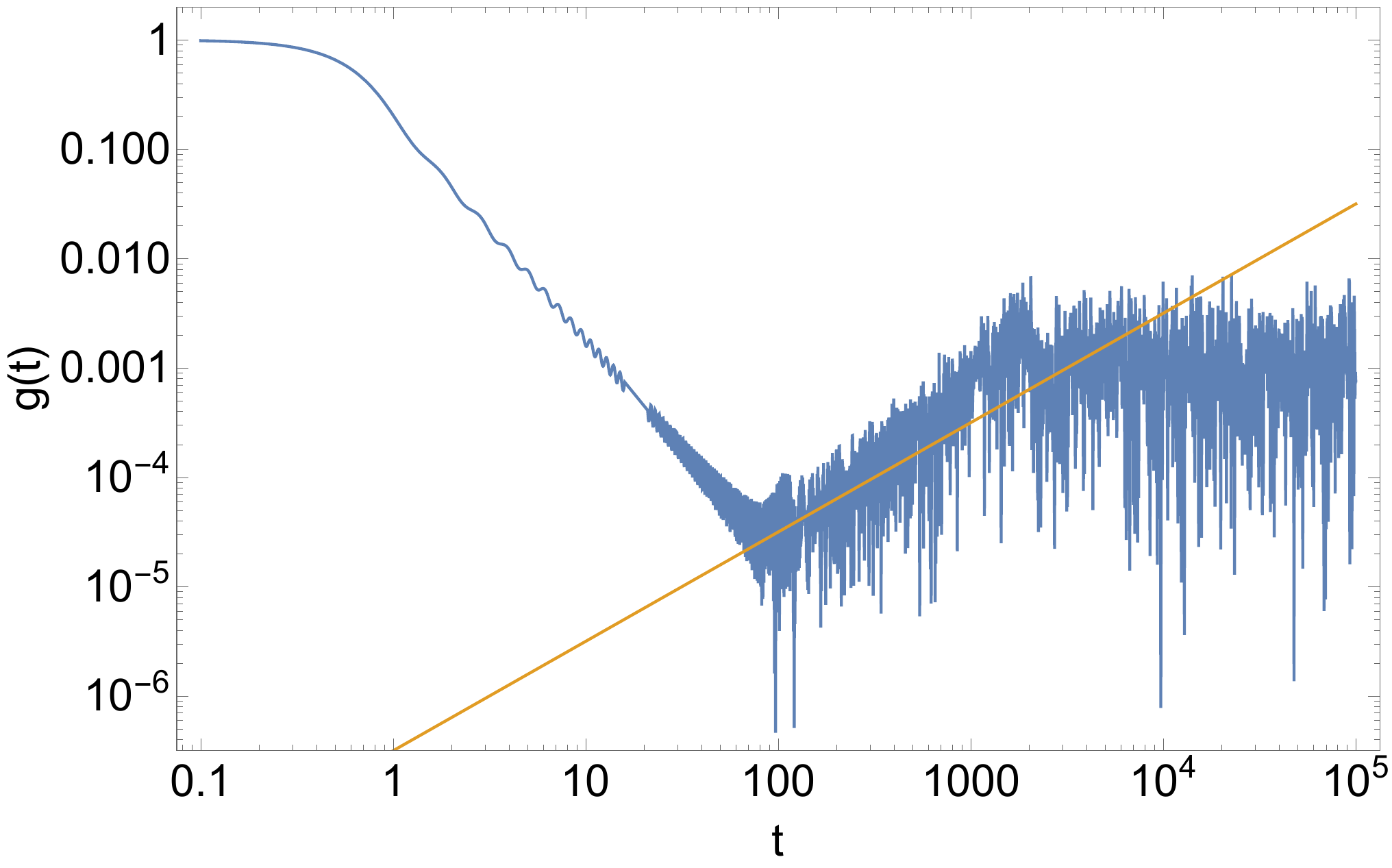}
         \caption{$\epsilon_0=10^{-5}$}
         \label{fig_1_1}
     \end{subfigure}
     \hfill
      \begin{subfigure}[b]{0.45\textwidth}
         \centering
         \includegraphics[width=\textwidth]{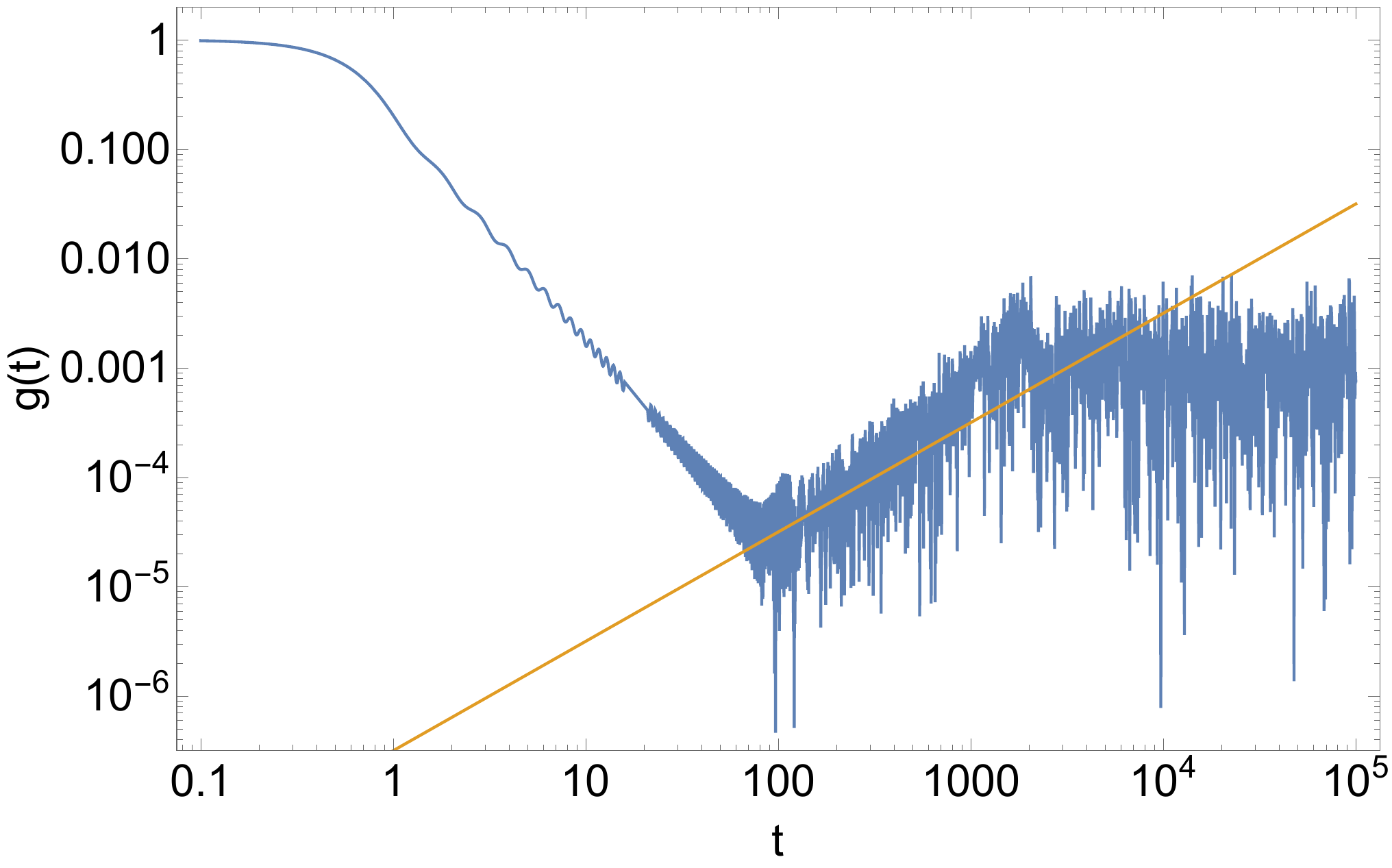}
         \caption{$\epsilon_0=10^{-7.5}$}
         \label{fig_1_05}
     \end{subfigure}
     \hfill
     \begin{subfigure}[b]{0.45\textwidth}
         \centering
         \includegraphics[width=\textwidth]{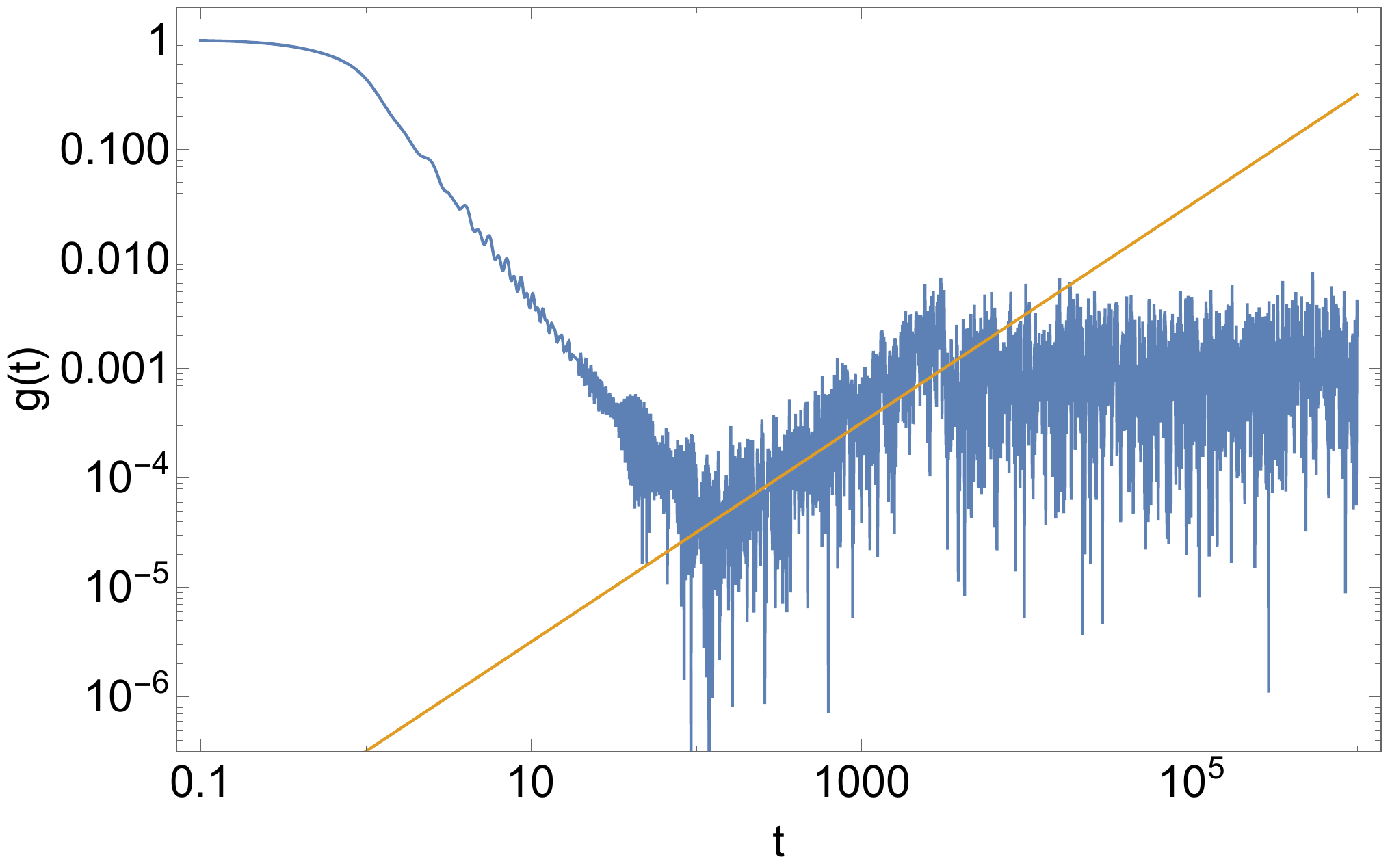}
         \caption{$\epsilon_0=10^{-8.5}$}
         \label{fig_1_005}
     \end{subfigure}
     \hfill
  \begin{subfigure}[b]{0.45\textwidth}
         \centering
         \includegraphics[width=\textwidth]{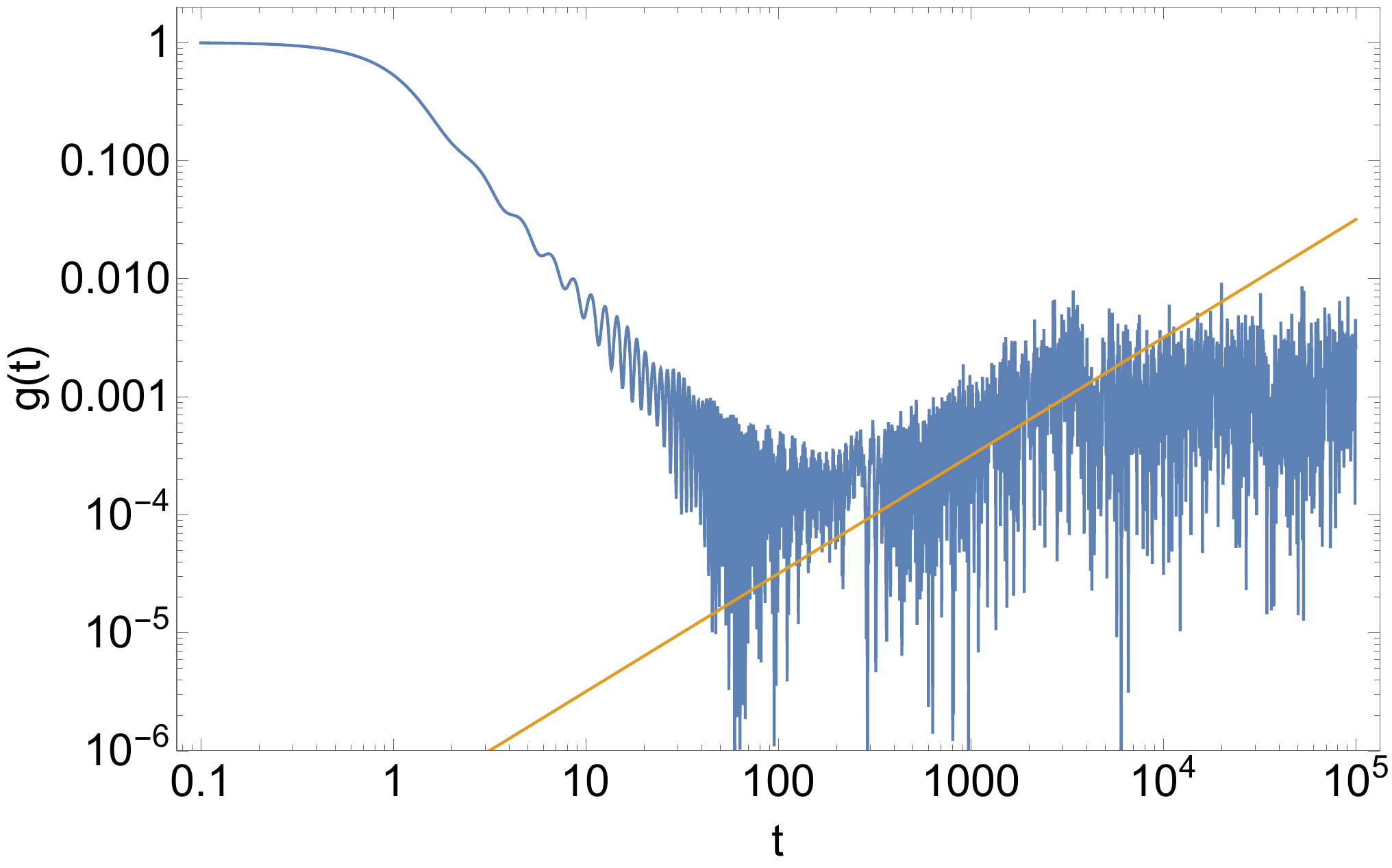}
         \caption{$\epsilon_0=10^{-9}$}
         \label{fig_1_0005}
     \end{subfigure}
        \caption{This set of figures shows how the slope of the ramp is approaching one as we move the brick wall close to the horizon. The yellow line has slope one. The noise at the end of the spectrum is caused by the removal of the first few roots from the spectrum. $n$ is fixed to one. Here $\beta=0$.}
        \label{WKB_sff}
\end{figure}

\begin{figure}
    \centering
    \includegraphics[width=.55\textwidth]{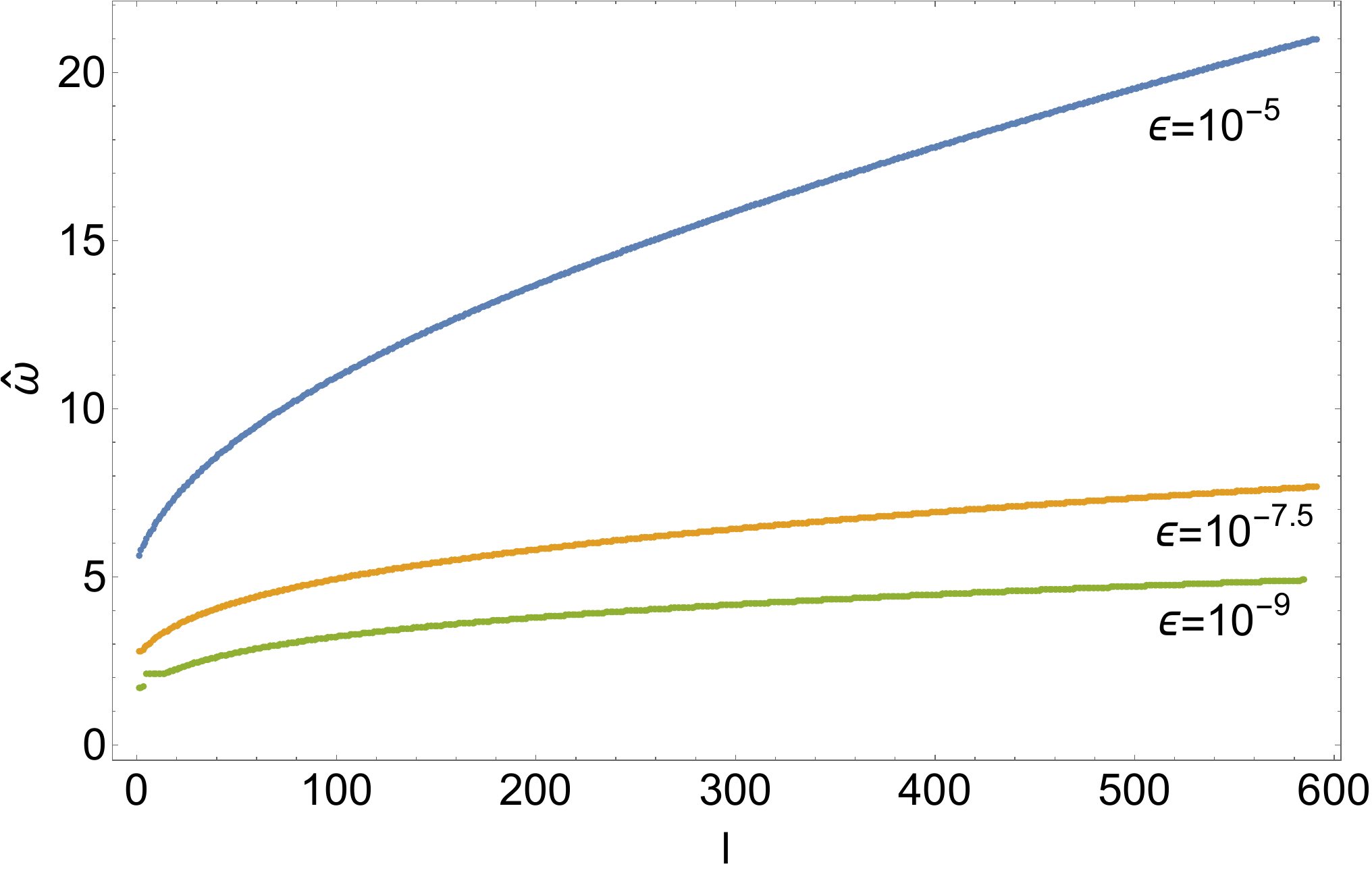}
   \caption{Behaviour of normal modes along the $l$ direction with varying $\epsilon_0$.}
    \label{spectrum_comparison_WKB}
\end{figure}

\begin{figure}
    \centering
    \includegraphics[width=.55\textwidth]{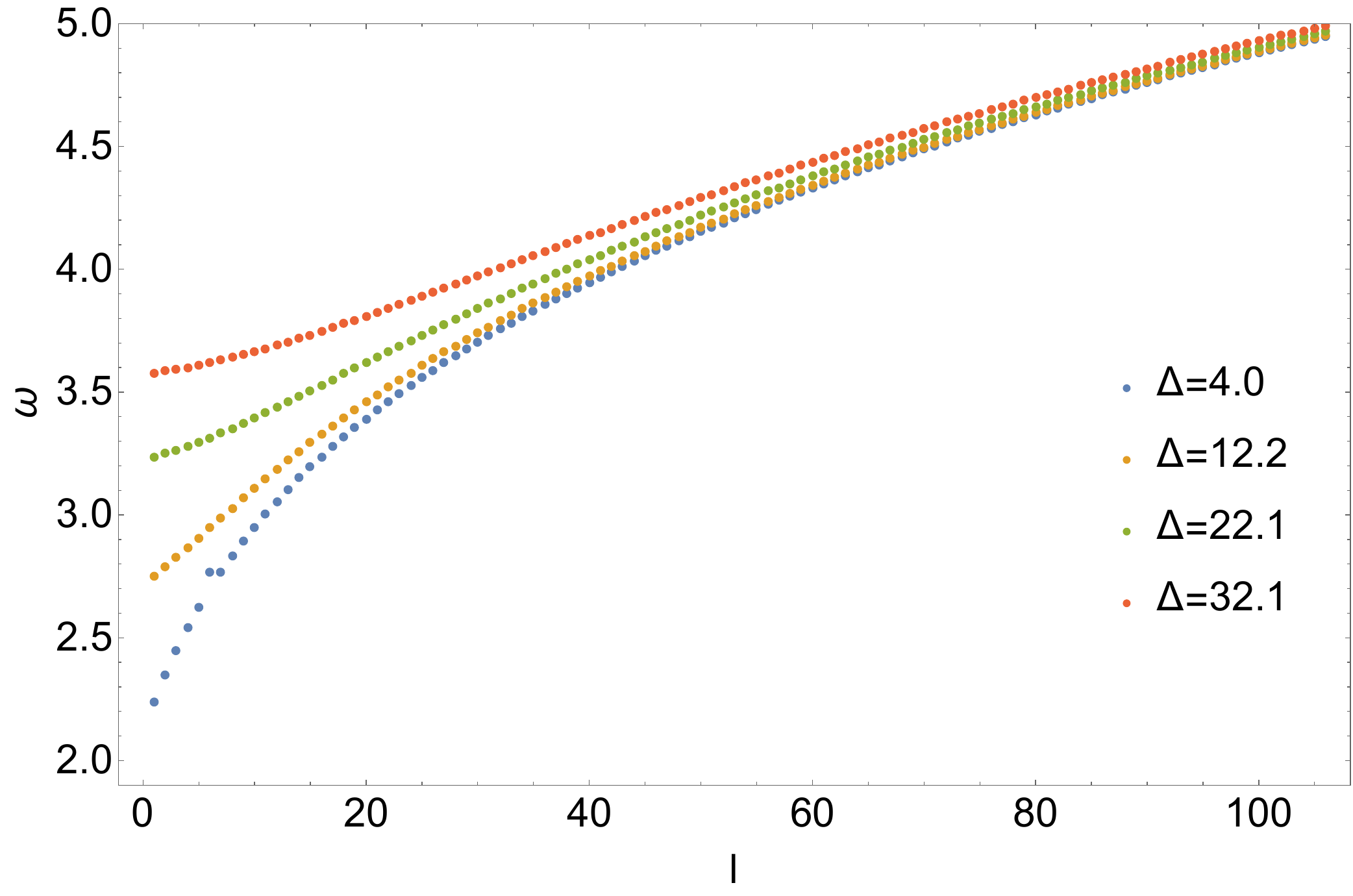}
   \caption{Behavior of the normal modes along $l$-direction with varying $\Delta$. This is opposite to the behavior observed using the Liouville CFT technique, but consistent with the results of BTZ black hole. Parameters : $r_H/l=1$, $\epsilon_0=10^{-7.5}$.}
    \label{delta_WKB}
\end{figure}

The near-horizon expansion of $V(r)$ is,
\begin{equation}\label{pot2}
    \lim_{\epsilon\to0} V(r=r_H+ \epsilon)=V_{\text{eff}}(\epsilon)=-\frac{A_2}{\epsilon^2}+\frac{A_1}{\epsilon }-A_0+O(\epsilon),
\end{equation}
where $r=r_H+\epsilon$, and,
\begin{eqnarray}
&& A_0 = \frac{7}{432} \left(12 \, l (l+2)-\omega^2-9\right) \ , \\
&& A_1 =\frac{18l (l+2)-5 \omega^2+117}{108 } \ , \\
&& A_2 =\frac{9+\omega^2}{36 } \ . 
\end{eqnarray}
\begin{figure}
    \centering
    \includegraphics[width=.55\textwidth]{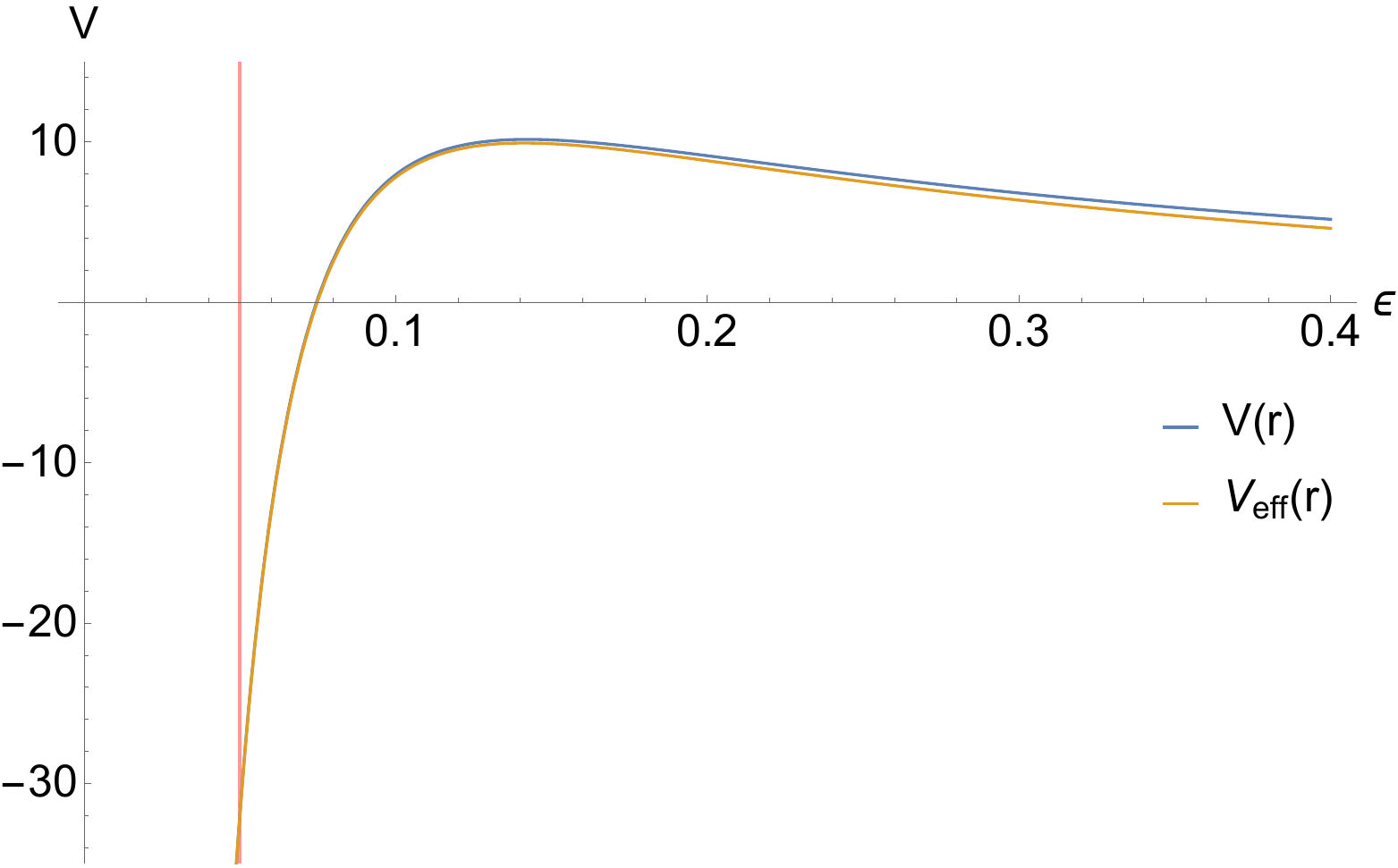}
   \caption{Comparison of \eqref{pot1} and \eqref{pot2}. This image shows that when the turning point is close to the horizon, we can approximate the WKB integration using  $V_{\text{eff}}$. Parameters: $l=3, \omega=0.3$, $r_H=1$.}
    \label{WKB_pot2}
\end{figure}
Under the above assumption, we can approximate the original $V(r)$ by \eqref{pot2}. A comparison of $V(\epsilon)$ and $V_\text{{eff}}(\epsilon)$ is shown in Figure \ref{WKB_pot2}. With this approximation, the WKB integration can be performed exactly, yielding the following result:
\begin{equation}\label{wkb_int}
    \int_{r_0}^{r_c} |V(r)|^{\frac{1}{2}} dr = -\sqrt{A_0 \, \epsilon_0 ^2-A_1 \, \epsilon_0 +A_2}-\frac{1}{2} \sqrt{A_2} \log \left(\frac{T_1+1}{T_1-1}\right)-\frac{A_1}{4 \sqrt{A_0}}   \log \left(\frac{T_2+1}{T_2-1}\right),
\end{equation}
where,
\begin{eqnarray}
&& T_1 = \frac{A_1 \, \epsilon_0-2 A_2}{2\sqrt{A_2(A_2-A_1 \, \epsilon_0+A_0 \, \epsilon_0^2)}} \, , \\
&& T_2 = \frac{-2 A_0 \, \epsilon_0+ A_1}{2\sqrt{A_0(A_2-A_1 \, \epsilon_0+A_0 \, \epsilon_0^2)}} \, .
\end{eqnarray}
Then according to WKB method,
\begin{equation}
    \int_{r_0}^{r_c} |V(r)|^{\frac{1}{2}} dr=\frac{3\pi}{4}+  n \pi ,
\end{equation}
where $n$ is the principal quantum number and $\epsilon_0=r_0-r_H$.  This equation cannot be solved analytically for $\omega_{n, l}$, so we have solved it numerically for different choices of $\epsilon_0$ using \textit{Mathematica}. The results are the following.\\
\begin{figure}
\begin{subfigure}{0.47\textwidth}
    \centering
    \includegraphics[width=\textwidth]{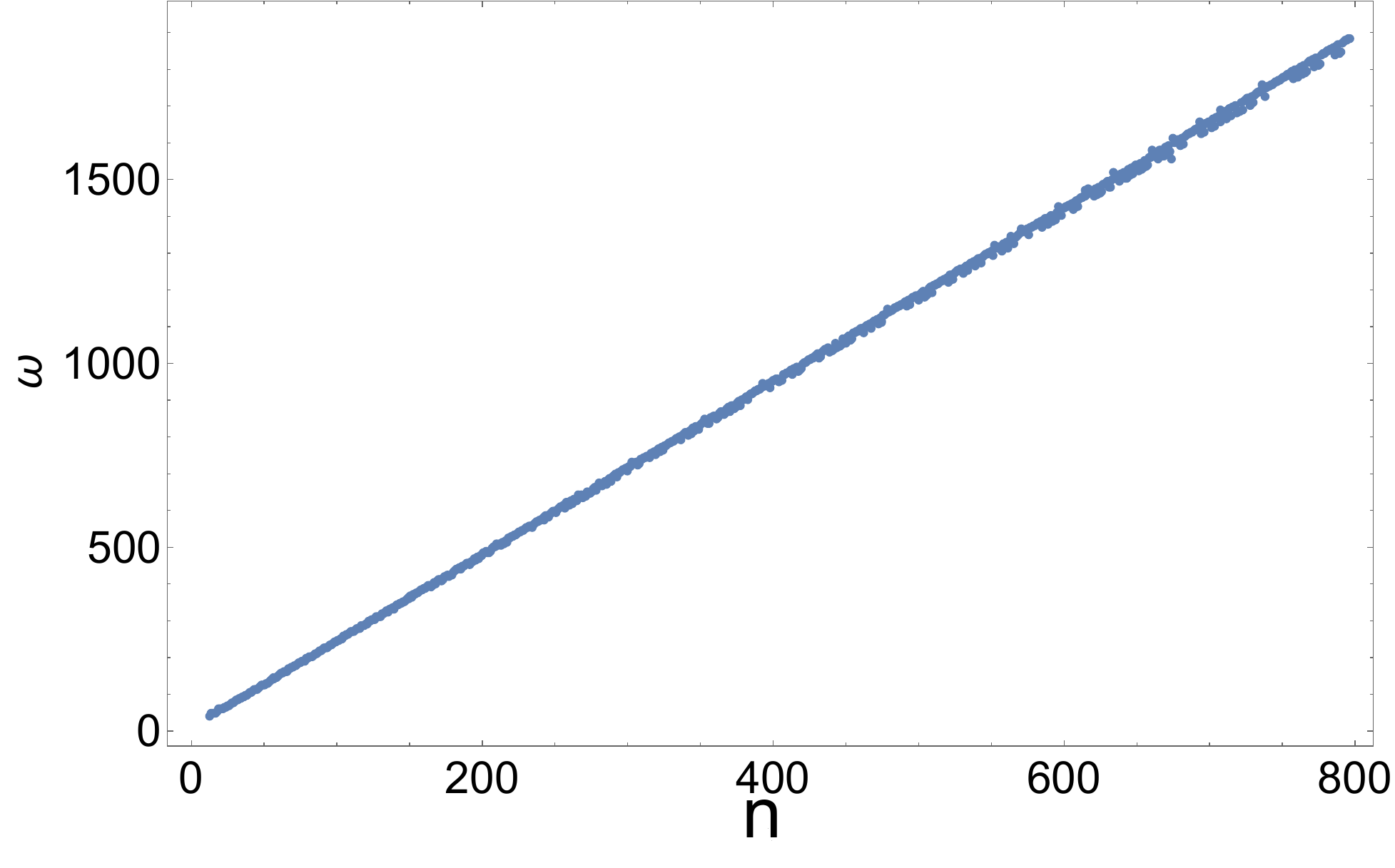}
    \end{subfigure}
    \hfill
    \begin{subfigure}{0.47\textwidth}
    \includegraphics[width=\textwidth]{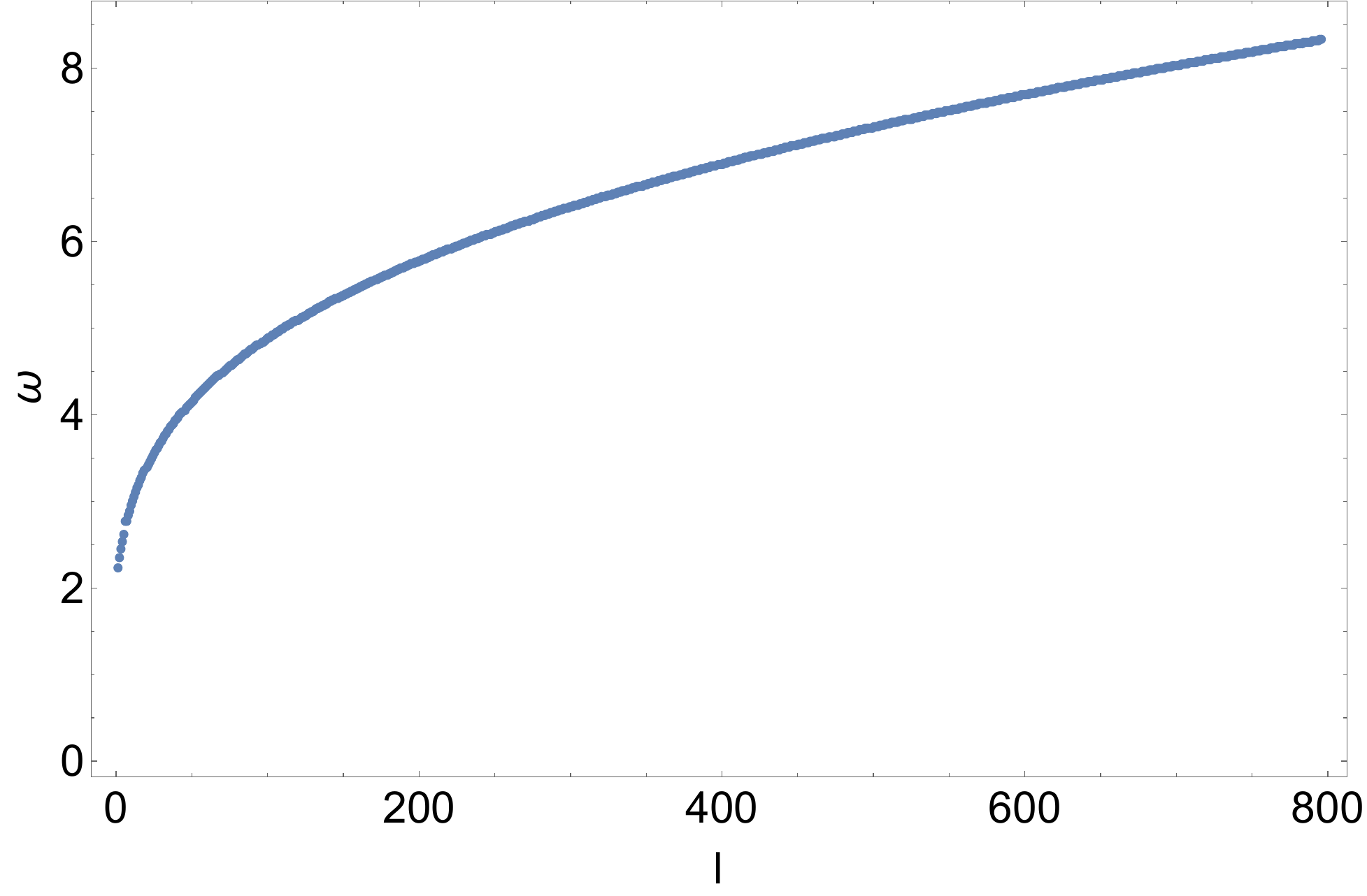}
    \end{subfigure}
    \caption{Spectrum along $n$ (left) and $l$ (right) direction using WKB method. Both $n$ and $l$ start form $10$, as low-lying modes are difficult to obtain with WKB.  Other parameters: $r_H=1$, $\epsilon_0=10^{-7.5}$. For the left figure, $l$ is fixed to $5$, whereas for the right figure, $n=1$.}
    \label{spectrum_WKB}
\end{figure}
In Figure \ref{spectrum_WKB}, we present the spectrum along the $n$ and $l$-directions. Note the striking similarity with the figures in \cite{Das:2022evy}. The spectrum along $n$ direction is linear, resembling that of a simple harmonic oscillator. Consequently, we do not observe any ramp structure in the SFF (see Figure \ref{sff_WKB} (left)). In contrast, the non-trivial functional dependence along $l$ direction leads to the presence of a ramp in the SFF along $l$ direction (see Figure \ref{sff_WKB} (right)). 
\begin{figure}
\begin{subfigure}{0.47\textwidth}
    \centering
    \includegraphics[width=\textwidth]{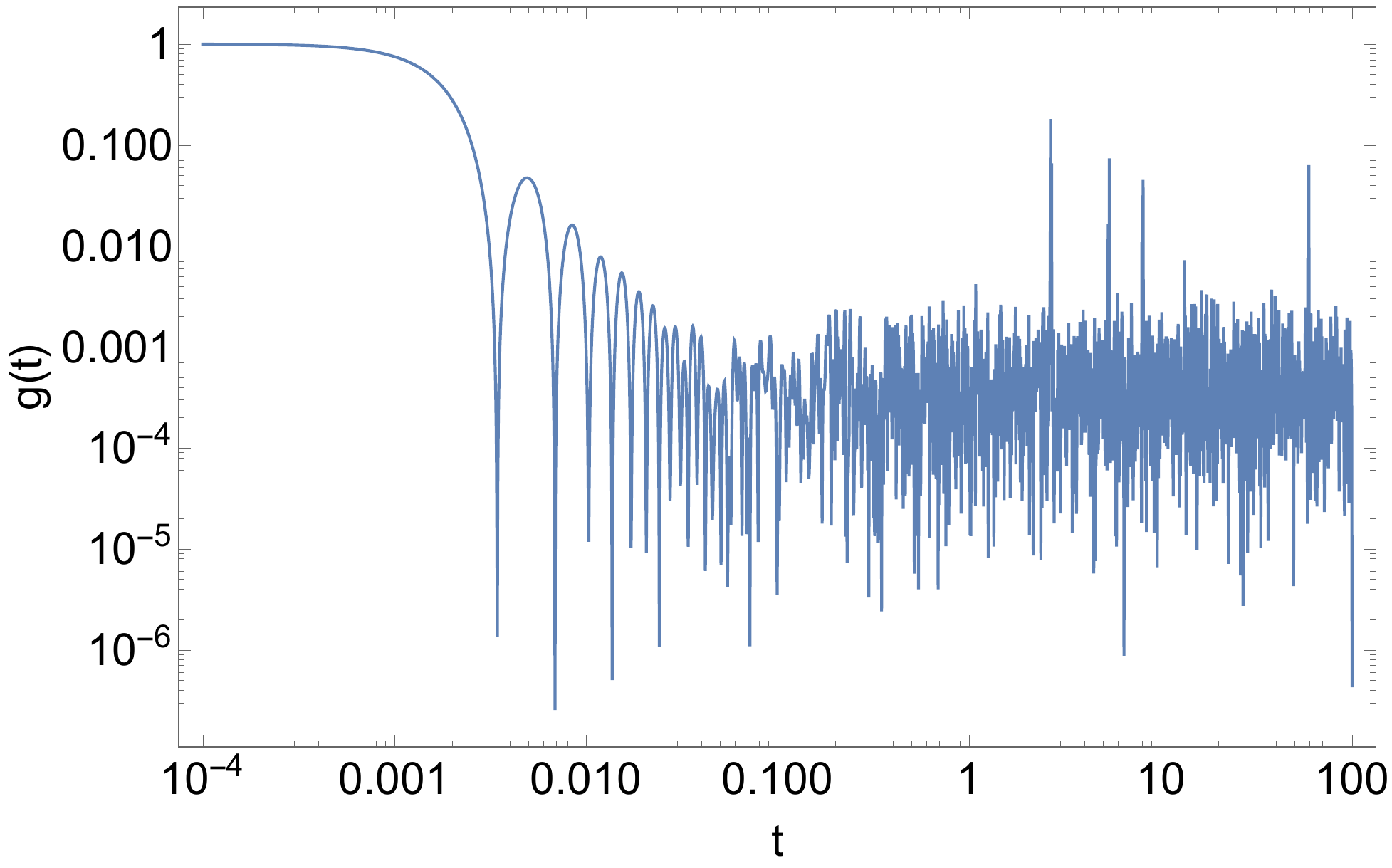}
    \end{subfigure}
    \hfill
    \begin{subfigure}{0.47\textwidth}
    \includegraphics[width=\textwidth]{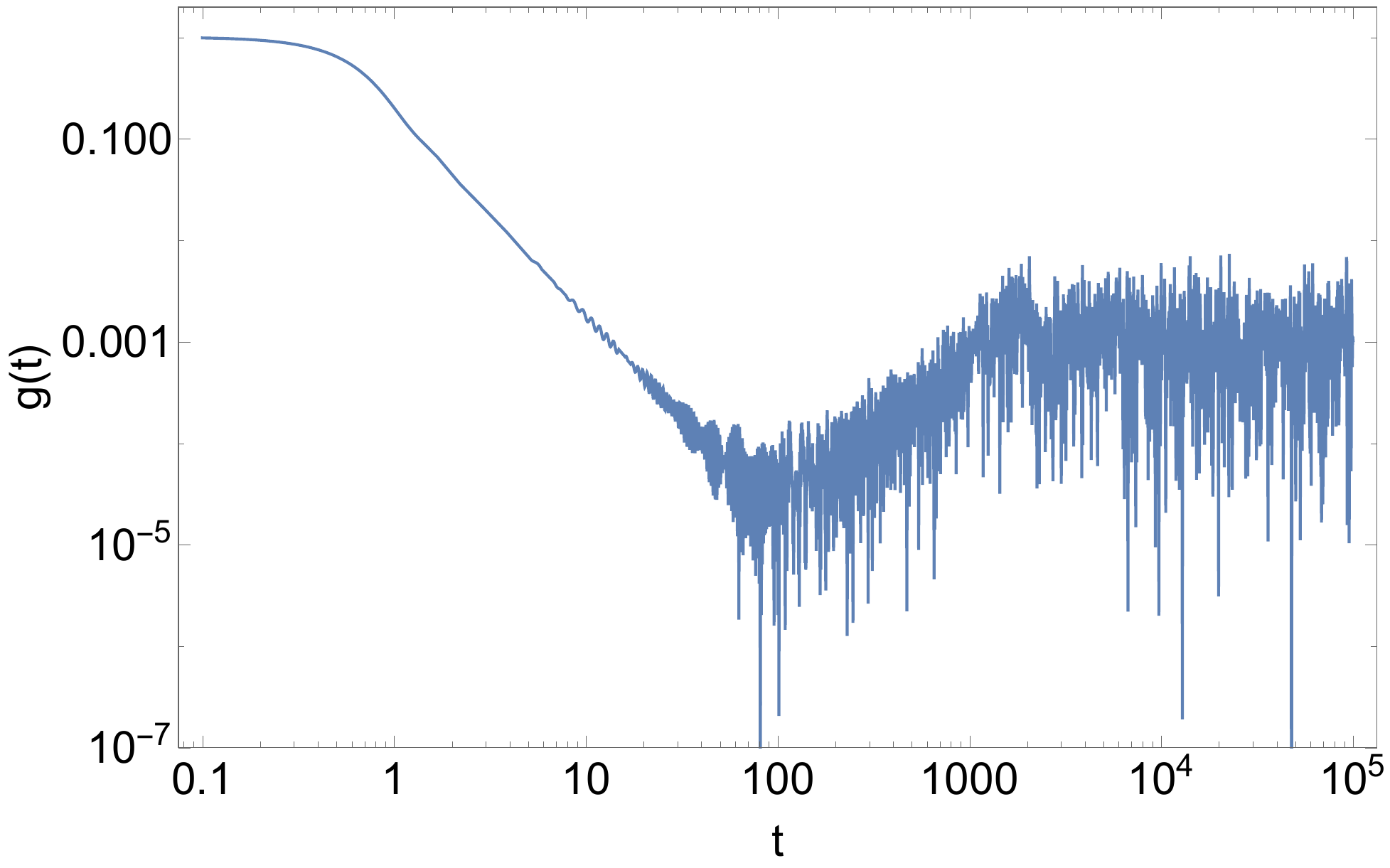}
    \end{subfigure}
    \caption{SFF along $n$ (left) and $l$(right) directions for the modes shown in Figure \ref{spectrum_WKB}. Here $\beta$ is fixed to zero.}
    \label{sff_WKB}
\end{figure}
Figure \ref{WKB_sff} shows the SFFs for different $\epsilon_0$. As $\epsilon_0$ decreases, the slope of the ramp approaches one, consistent with our previous observation of a linear ramp in  $2+1$ dimensions.
In Figure \ref{spectrum_comparison_WKB}, we illustrate the spectrum for different values of $\epsilon_0$. As $\epsilon_0$ decreases, the $\omega$ values become increasingly closer to each other. This quasi-degenerate property leads to the ramp observed in the Spectral Form Factor (SFF). Due to the instability of low-lying modes, as shown in Figure \ref{spectrum_comparison_WKB}, our numerical method cannot handle values of $\epsilon_0$ below $10^{-9}$.\\
Finally, in Figure \ref{delta_WKB}, we present the $\Delta$-dependence of the modes along the $l$-direction. This behavior is the opposite of what was obtained using the Liouville CFT technique, as shown in Figure \ref{delta_dep} (left).
\newpage

\section{Solving Heun equation perturbatively}\label{sec:5}

In this section, we will solve the radial Heun equation using a perturbative method, as discussed in \cite{Musiri:2003rv}. To begin, we rewrite the radial equation \eqref{eom2} in terms of a dimensionless coordinate $y=\frac{r_H^2}{r^2}$, resulting in:
\begin{equation}\label{eom4}
    y^3(1-y^2) \frac{d}{dy} \left( \frac{1-y^2}{y} \phi'(y)   \right)+ \left( \frac{\hat{\omega}^2}{4} y- \frac{\hat{l}^2}{4} y(1-y^2)  -\frac{\hat{\mu}^2}{4} (1-y^2)  \right) \phi(y)=0,
\end{equation}
where we have defined $\hat{\omega}=\omega/r_H$, $\hat{l}^2=\frac{l(l+2)}{r_H^2}$ and $\hat{\mu}= \mu$. We choose the following ansatz for $\phi(y)$,
\begin{equation}
    \phi(y)=y^2(1-y)^{\frac{-i \hat{\omega}}{4}} \left(\frac{1+y}{2}\right)^{-\frac{\hat{\omega}}{4}} F(y)\, .
\end{equation}
Substituting this into \eqref{eom4}, we obtain:
\begin{equation}\label{eom5}
    F''(y)+\left( \frac{3}{y}+\frac{1-i \hat{\omega}/2}{y-1} +\frac{1-\hat{\omega}/2}{y+1}  \right) F'(y)+ \frac{(2-(1+i)\hat{\omega}/4)^2 y-q}{y(y^2-1)}F(y)=0\, ,
\end{equation}
where
\begin{equation}\label{qeqn}
    q=\frac{3(i-1)}{4} \hat{\omega}- \frac{\hat{l}^2}{4}+\frac{\hat{\omega}^2}{4}\, .
\end{equation}
Introducing another new variable $x=y^2$,  \eqref{eom5} can be rewritten as:
\begin{align}\label{eom6}
    \mathcal{H}F(x)&=x(1-x) F''(x)+\frac{1}{4} \left((1+i) x (\omega +(-6+6 i))-(1-i) \sqrt{x} \omega +8\right) F'(x) \nonumber \\ & \hspace{4cm}+\frac{1}{32} \left(\frac{8 q}{\sqrt{x}}-i (\omega +(-4+4 i))^2\right) F(x)=0.
\end{align}
To solve the equation perturbatively, we decompose the Hamiltonian into two terms ($\mathcal{H}_0 + \mathcal{H}_1$) in such a way that $\mathcal{H}_1$ can be treated as a perturbation in $\hat{\omega}$:

\begin{align}
    \mathcal{H}_0 &=x(1-x) \frac{d^2}{dx^2}+\left( 2-\frac{1-i}{4}\hat{\omega}-(3-\frac{1+i}{4}\hat{\omega})x \right) \frac{d}{dx}-\frac{1}{4}\left((2-\frac{1+i}{4}\hat{\omega})^2-q) \right), \nonumber\\
    \mathcal{H}_1 &=(1-\sqrt{x})\left( \frac{1-i}{4}\hat{\omega} \frac{d}{dx}+\frac{q}{4\sqrt{x}} \right).
\end{align}
The above decomposition has an additional benefit: $\mathcal{H}_0$ is a hypergeometric differential equation. Furthermore, we expand $F(x)$ as a perturbative series in $\hat{\omega}$.:
\begin{equation}
    F(x)=F_0(x)+F_1(x)+F_2(x)\ldots
\end{equation}
This expansion allows us to rewrite \eqref{eom6} formally as:
\begin{align}
    &(\mathcal{H}_0+\mathcal{H}_1)(F_0+F_1+\ldots)=0 \nonumber\\
    \implies &\mathcal{H}_0 F_0+(\mathcal{H}_0 F_1+\mathcal{H}_1 F_0)+(H_0 F_2+H_1 F_1)+\ldots=0. \nonumber
\end{align}
At zeroth order, we have:
\begin{equation*}
    \mathcal{H}_0 F_0=0 .
\end{equation*}
At first order:
\begin{align*}
    &\mathcal{H}_0 F_1 +\mathcal{H}_1 F_0 =0\\
    \implies &F_1=-\mathcal{H}_0^{-1} \mathcal{H}_1 F_0=-\mathcal{D} F_0 .\\
\end{align*}
In second order:
\begin{align}
    &\mathcal{H}_0 F_2+\mathcal{H}_1 F_1=0 \\
    \implies F_2 &=\mathcal{H}_0^{-1} \mathcal{H}_1 \mathcal{H}_0^{-1} \mathcal{H}_1 F_0\\
    &=\mathcal{D}^2 F_0.
\end{align}
Similarly, at the $n$-th order, we have, $F_n=(-1)^n \mathcal{D}^n F_0$. Thus, once $F_0$ is known, we can in principle construct the full perturbative series for $F(x)$.

\begin{figure}
\begin{subfigure}{0.47\textwidth}
    \centering
    \includegraphics[width=\textwidth]{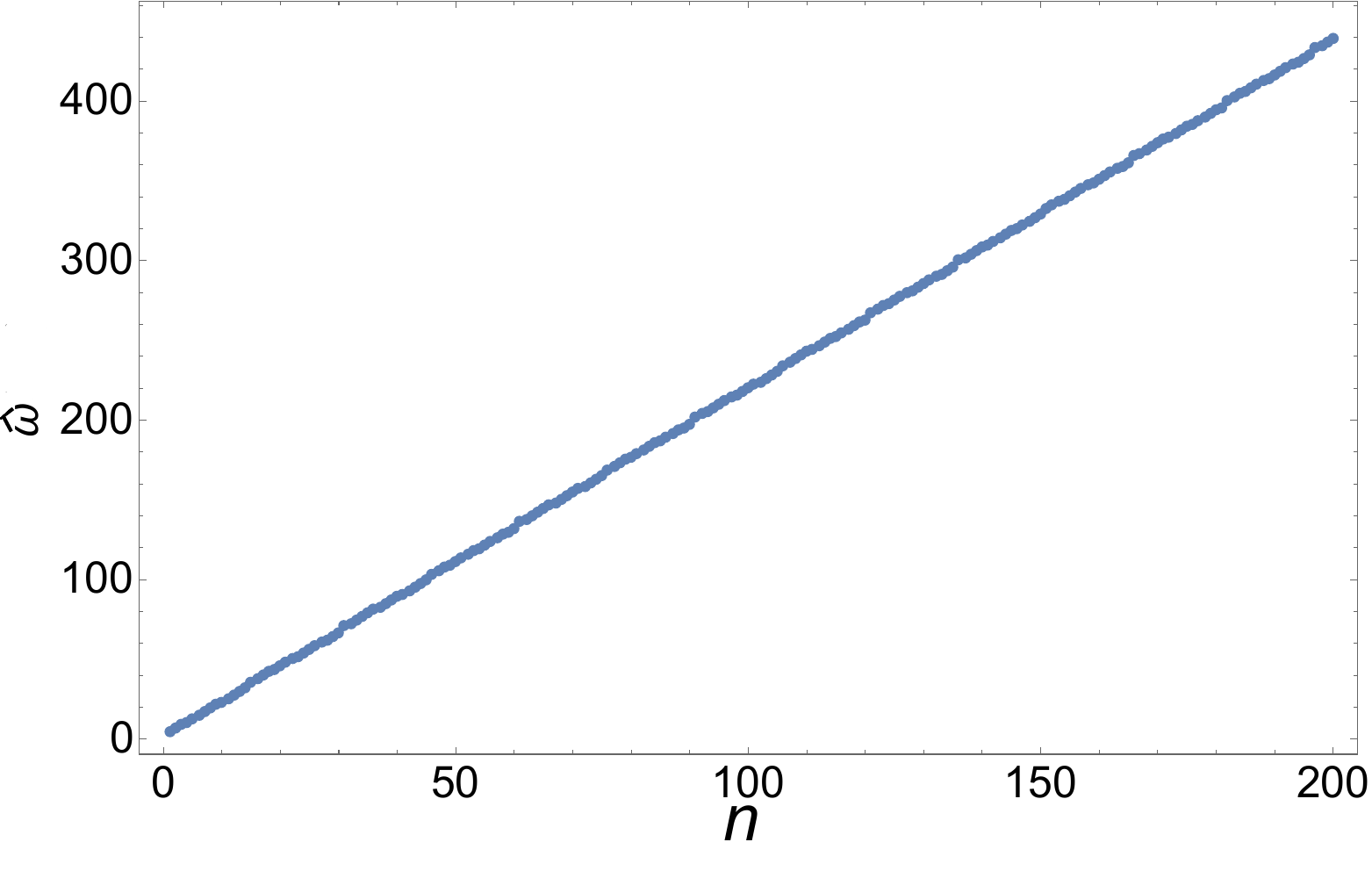}
    \end{subfigure}
    \hfill
    \begin{subfigure}{0.47\textwidth}
    \includegraphics[width=\textwidth]{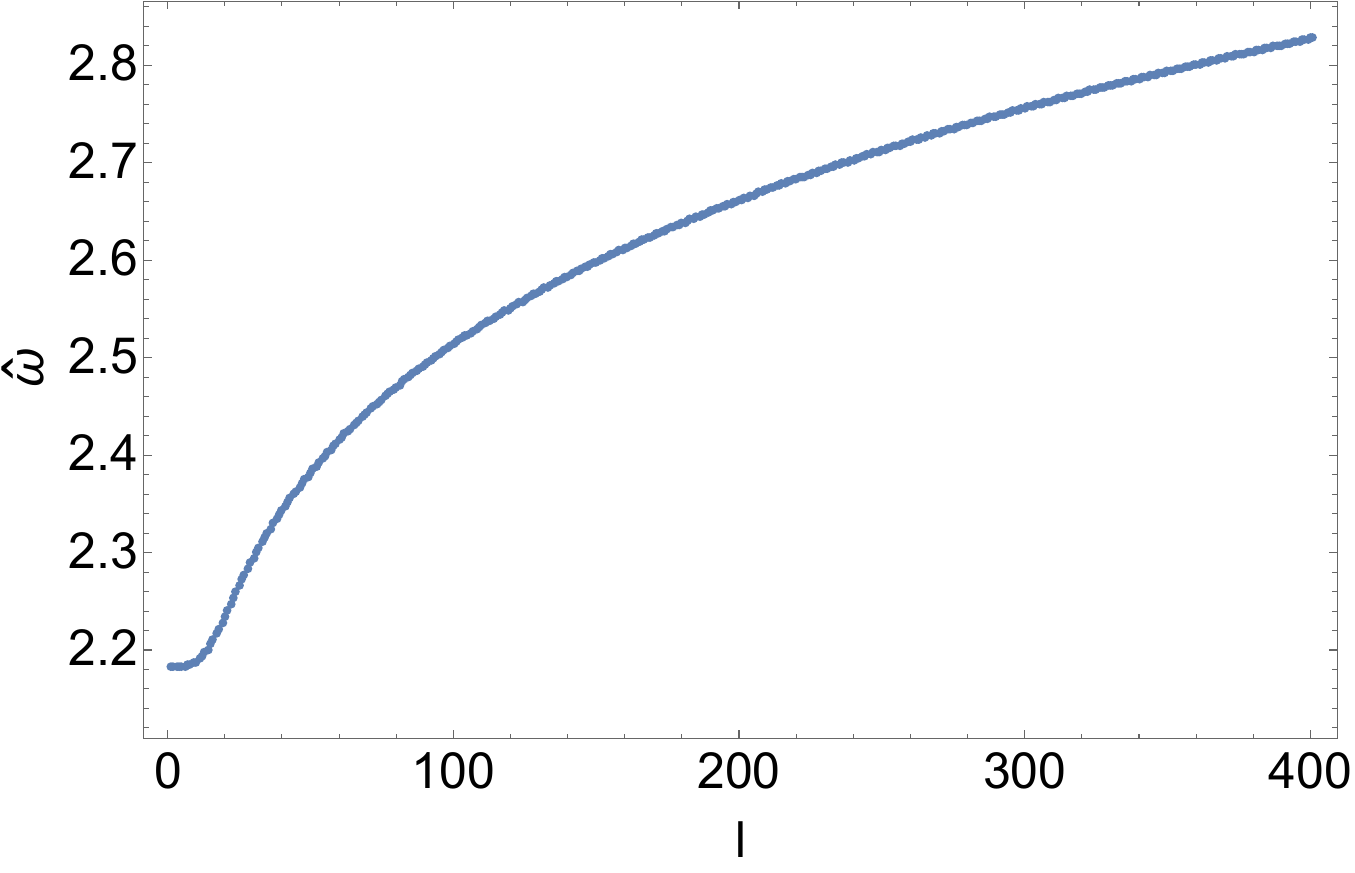}
    \end{subfigure}
    \caption{Spectrum along $n$ (left) and $l$ (right) direction for $r_H=10$.}
    \label{spectrum_el_n}
\end{figure}

\begin{figure}
    \centering
    \includegraphics[width=.55\textwidth]{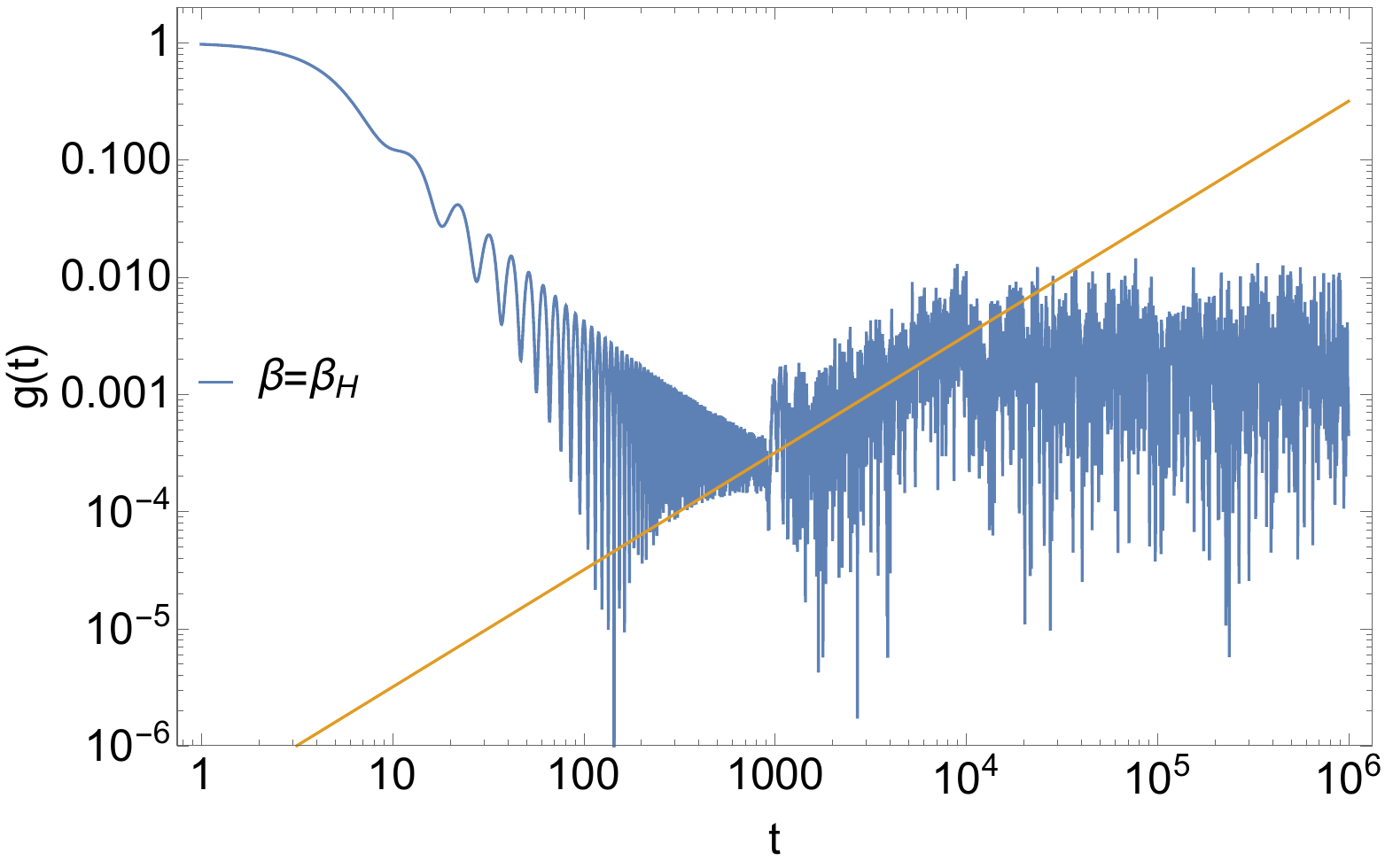}
   \caption{SFF for the normal modes along $l$ direction with fixed $n=0$. Here $l_{cut}=400$ and $\beta=\beta_H$}
    \label{SFF_pertur}
\end{figure}

As already mentioned earlier, at zeroth order, the radial equation reduces to a hypergeometric equation:
\begin{align}\label{hyp1}
     x(1-x)F_0''(x)+\left( c-(1+a+b)x \right) F_0'(x)- a\, b F_0(x)=0\, ,
\end{align}
where,
\begin{equation}
    a,b=1+\frac{1}{2} \left(-\frac{1+i}{4} \hat{\omega}\pm \sqrt{q}  \right), \hspace{0.5cm} c=2-\frac{1-i}{4} \hat{\omega}.
\end{equation}
The solution of \eqref{hyp1} is given in terms of hypergeometric functions:
\begin{equation}\label{sol1}
    F_0(x)=c_1 \, _{2}F_1(a, b, c, x)+c_2 \, x^{1-c}\, _{2}F_1(1+a-c, 1+b-c, 2-c, x)\, .
\end{equation}
The near-boundary expansion of \eqref{sol1} is:
\begin{equation*}
    F_0(x) \sim c_1+c_2 x^{1-c}\, ,
\end{equation*}
which implies: 
\begin{equation*}
    \psi \sim (1-y)^{-\frac{i \hat{\omega}}{4}} \left(\frac{1+y}{2}  \right)^{-\frac{\hat{\omega}}{4}} (c_1  \, y^2+c_2 \, y^{\frac{1-i}{2}\hat{\omega}})\, .
\end{equation*}
The normalizable boundary condition ($\psi \rightarrow 0$) implies $c_2=0$. \\
The near-horizon behavior of \eqref{sol1} is given by:
\begin{equation}\label{gr7}
    F_0(x) \sim A+B \, (1-x)^{\frac{i \hat{\omega}}{2}}\, ,
\end{equation}
where,
\begin{align} \label{gr8}
    A &= \frac{\Gamma(c)\Gamma(c-a-b)}{\Gamma(c-a)\Gamma(c-b)}, \nonumber \\ 
    B &= \frac{\Gamma(c)\Gamma(a+b-c)}{\Gamma(a)\Gamma(b)}\, .
\end{align}
If we impose the Dirichlet boundary condition at the stretched horizon (close to the event horizon):
\begin{align}
    &F_0(x_0) \sim A+B \, (1-x_0)^{\frac{i \hat{\omega}}{2}}=0  \nonumber \\
    &\implies (1-x_0)^{\frac{i \hat{\omega}}{2}}=-\frac{A}{B}  \nonumber \\
    &\implies\epsilon ^{\frac{i \hat{\omega}}{2}}=-\frac{A}{B} \nonumber \\
    &\implies \text{Arg} \bigg[  \frac{\Gamma(a+b-c) \Gamma(c-a) \Gamma(c-b)}{\Gamma(c-a-b) \Gamma(a) \Gamma(b)} \bigg]+\frac{\hat{\omega}}{2} \log \epsilon=(2 n+1)\pi \, .
\end{align}
where, $n \in \mathbf{Z}$. Solving this equation in `Mathematica', we obtain $\omega$ as a function of $n$ and $l$. Figure [\ref{spectrum_el_n}] shows the behavior of the spectrum in the $n$ and $l$ directions. As we can see again, the spectrum has a linear behavior along $n$ and a non-linear behavior along $l$. 
With the spectrum in hand, we can now compute the SFF. Figure [\ref{SFF_pertur}] shows SFF along the $l$ direction at temperature $\beta_H$ which is the Hawking temperature of the black hole, but it is worth mentioning that $\beta$ is nothing but a parameter here.

So far, we have discussed the zeroth-order solution. However, one can find the higher-order corrections to the normal modes by adding order-by-order corrections to the wave function. Explaining all the technical details is beyond the scope of this paper. However, a thorough discussion of this topic can be found in \cite{Musiri:2003rv}.

\newpage

\section{Green's function and emerging thermality}\label{sec:6}

\subsection{In momentum space}
In this section, we closely examine the analytic structure of the Green's function, following the approaches of \cite{Giusto:2023awo} and \cite{Banerjee:2024dpl}. For any asymptotic AdS geometry, the Son-Starinets \cite{Son:2002sd} prescription provides a method to compute the Green's function of the boundary theory as the ratio of the normalizable and non-normalizable modes. From \eqref{eq00}, this implies:
\begin{equation}
    G=\frac{b_{11}+ R_{c_1c_2} a_{11}}{b_{22}+R_{c_1c_2} a_{22}}\ .
\end{equation}

\begin{figure}
\begin{subfigure}{0.47\textwidth}
    \centering
    \includegraphics[width=\textwidth]{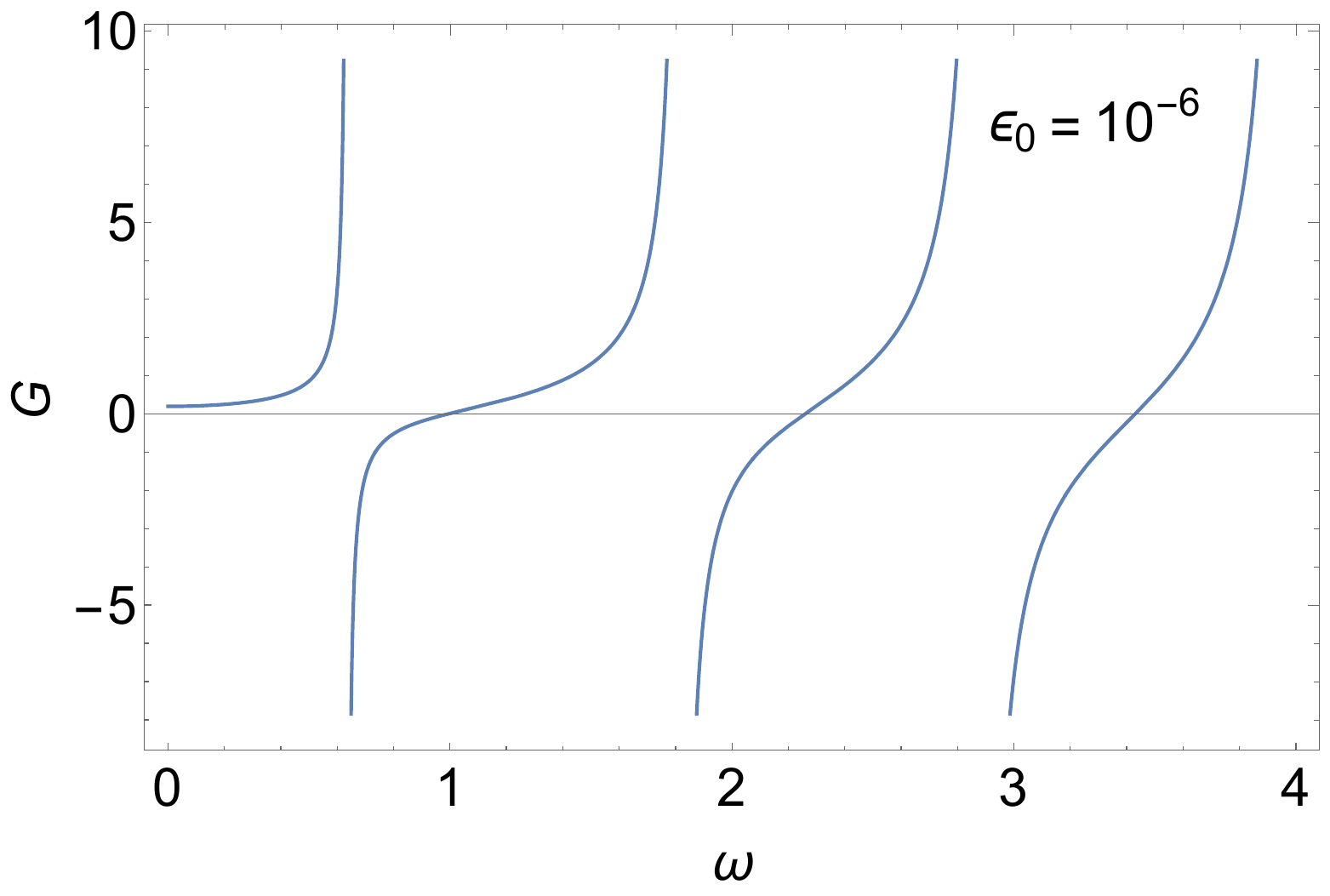}
    \end{subfigure}
    \hfill
    \begin{subfigure}{0.47\textwidth}
    \includegraphics[width=\textwidth]{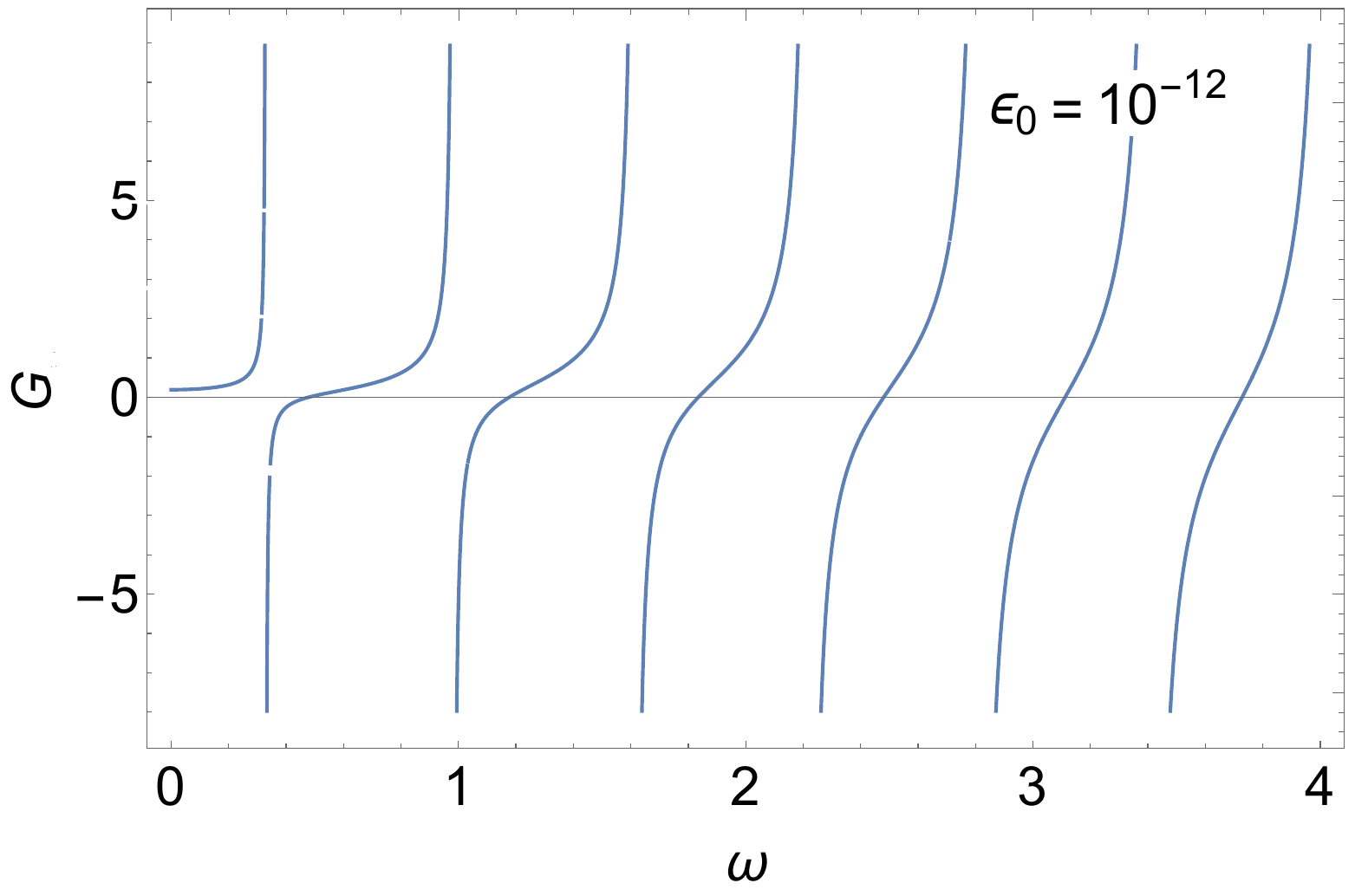}
    \end{subfigure}
    \caption{Plot of Green's function, $G(\omega, l=1)$ for two different $\epsilon_0$. The main point to note is that as $\epsilon_0$ decreases the poles start to accumulate on the real line.}
    \label{green_pole}
\end{figure}

The poles of this Green's function correspond to the excitations of the system when perturbed by the boundary operator dual to the scalar field in the bulk. Since the system is in a pure state (with no infalling boundary condition), we expect the poles to lie on the real line. Except that, the function is analytic in the whole complex plane. In Figure \ref{green_pole}, we plot the Green's function $G$ as a function of $\omega$. As $z_0$ is decreased, i.e., as the brick wall approaches the horizon, the poles come closer together. In the limit $\epsilon_0\rightarrow0$, these poles are so close that we can approximate them as a branch cut. However, there is a caveat: for any given $\epsilon_0$, there is always a maximum gap $\Delta \omega$ (noting that the spectrum is not equidistant), which sets a minimum timescale of $\sim \frac{1}{\Delta \omega}$. Beyond this timescale, a boundary observer begins to probe the discreteness of the spectrum.

In the position space correlator, the implication of this pole condensation is very interesting. To see this, let's first rewrite $G$ as follows:
\begin{align}\label{gre1}
    G&=\frac{b_{11}+ R_{c_1c_2} a_{11}}{b_{22}+R_{c_1c_2} a_{22}} \nonumber \\
    &= \frac{\frac{b_{11}}{b_{22}}+R_{c_1c_2} \frac{a_{11}}{b_{22}}}{1+R_{c_1c_2} \frac{a_{22}}{b_{22}}} \nonumber \\
    &= \frac{G_{R}^{BH}+R_{c_1c_2} \frac{a_{11}}{b_{22}}}{1+R_{c_1c_2} \frac{a_{22}}{b_{22}}}\ ,
\end{align}
where the retarded correlator of the black hole, $G_{R}^{BH}$, is given in \eqref{greenfn}. The poles of the Green's function occur when the denominator of \eqref{gre1} goes to zero, which is nothing but the quantization equation \eqref{quan0}. This happens because the condition: denominator$=0$ corresponds to the vanishing of the non-normalizable mode, which is exactly the quantization condition (remember that the Dirichlet condition is already imposed in writing \eqref{eq00}).\\
The residue at these simple poles (labeled by $\omega_k$) is given by,
\begin{equation}
    \text{Res}(G, \omega_k)= \frac{\left(G_R^{BH}+  R_{c_1c_2} \frac{a_{11}}{b_{22}} \right)\bigg\rvert_{\omega=\omega_k}}{\partial_{\omega} \left( 1+ R_{c_1c_2} \frac{a_{22}}{b_{22}} \right)\bigg\rvert_{\omega=\omega_k}} \ .
\end{equation}
Now at $\omega=\omega_k$, denominator of \eqref{gre1} is zero i.e. $R_{c_1c_2}=-\frac{b_{22}}{b_{11}}$ which implies numerator can be written as,
\begin{align}
    \left(G_R^{BH}+  R_{c_1c_2} \frac{a_{11}}{b_{22}} \right)\bigg\rvert_{\omega=\omega_k} &= G_R^{BH}-\frac{a_{11}}{a_{22}} \bigg\rvert_{\omega=\omega_k} \nonumber \\
    &= G_R^{BH}-(G_R^{BH})^*\bigg\rvert_{\omega=\omega_k} \nonumber \\
    &= 2i \, \text{Im} G_R^{BH}(\omega, l) \bigg\rvert_{\omega=\omega_k}\ ,
\end{align}
whereas denominator becomes
\begin{align}
    \partial_{\omega} \left( 1+ R_{c_1c_2} \frac{a_{22}}{b_{22}} \right)\bigg\rvert_{\omega=\omega_k} &= \partial_{\omega} e^{-i \theta(\omega, l)}\bigg\rvert_{\omega=\omega_k}\ ,
\end{align}
where $\theta(\omega, l)$ is given in a footnote after \eqref{quan0}. Thus, the residue is,
\begin{equation}
    \text{Res}(G, \omega_k)= 2i \frac{\text{Im} G_R^{BH} (\omega, l) \bigg\rvert_{\omega=\omega_k}}{\partial_{\omega} e^{-i \theta(\omega, l)}\bigg\rvert_{\omega=\omega_k}}\ .
\end{equation}
If a function has simple poles at $\omega_1, \omega_2, \ldots, \omega_k, \ldots$, it can be expressed as,
\begin{equation}
    G(\omega, l)= \sum_{k} \left(\frac{\omega^2}{\omega_k^2}  \right)^{\delta}\frac{\text{Res}(G, \omega_k)}{\omega-\omega_k}.
\end{equation}
The additional factor is included to ensure the convergence of $G$ on the contour as $\omega \rightarrow \infty$. When the poles are very closely spaced, the sum can be approximated by an integral, given by the following expression:
\begin{equation}\label{rho}
     G(\omega, l)=\int d\omega_k  \frac{\rho_{\omega}(\omega_k, l)}{\omega-\omega_k} \quad \text{with} \, \rho_{\omega}(\omega_k, l)= \frac{dk}{d\omega_k}  \left(\frac{\omega^2}{\omega_k^2}  \right)^{\delta} \text{Res}(G, \omega_k)\ ,
\end{equation}
which implies
\begin{align}\label{branch0}
    G(\omega+i\epsilon, l)-G(\omega-i\epsilon, l) &\approx \int d\omega_k \rho_{\omega}(\omega_k, l) 2 \pi i \delta(\omega-\omega_k) \nonumber \\
    &= 2\pi i \rho_{\omega}(\omega, l)\ .
\end{align}
Thus, when the poles are densely packed, they can be approximated by a branch cut, with the discontinuity given by \eqref{branch0}.\\

We observe a similar feature as the brick wall approaches the event horizon, allowing us to make analogous approximations for the correlator in \eqref{gre1}. To compute the discontinuity, we evaluate:
\begin{align}
    \frac{dk}{d \omega_k} &= \frac{1}{2\pi} \frac{d\theta(\omega_k, l)}{d\omega_k} \nonumber \\
    & = -\frac{1}{2\pi i} e^{i \theta(\omega_k, l)} \partial_{\omega_k} e^{-i \theta(\omega_k, l)} \nonumber \\
    &= -\frac{1}{2\pi i} \partial_{\omega_k} e^{-i \theta(\omega_k, l)}\ .
\end{align}
Substituting this into the expression in \eqref{rho}, we obtain,
\begin{equation}\label{eqrho1}
     \rho_{\omega}(\omega_k, l)= -\frac{1}{\pi}  \left(\frac{\omega^2}{\omega_k^2}  \right)^{\Delta} \text{Im} G_R^{BH}(\omega, l_0) \bigg\rvert_{\omega=\omega_k}\ .
\end{equation}
Given the correlator in \eqref{gre1}, we can define the retarded, Feynman, and Wightman correlators as follows:
\begin{eqnarray}
    G_R(\omega, l)=G(\omega+i\epsilon, l)=\sum_{k} \left(\frac{\omega^2}{\omega_k^2}  \right)^{\Delta}\frac{\text{Res}(G, \omega_k)}{\omega-\omega_k+i\epsilon} \ ,
\end{eqnarray}
\begin{equation}
    G_F(\omega, l)=\sum_{\omega_k>0} \left(\frac{\omega^2}{\omega_k^2}  \right)^{\Delta}\frac{\text{Res}(G, \omega_k)}{\omega-\omega_k+i\epsilon}+\sum_{\omega_k<0} \left(\frac{\omega^2}{\omega_k^2}  \right)^{\Delta}\frac{\text{Res}(G, \omega_k)}{\omega-\omega_k-i\epsilon} \ ,
\end{equation}
\begin{eqnarray}
    G_W(\omega, l)=- \text{sign}\omega \, \text{Im}G_R(\omega, l)\ .
\end{eqnarray}
Thus, the imaginary part of the retarded Green's function can be expressed as:
\begin{equation}\label{eqrho2}
    \text{Im}G_R(\omega, l)=\frac{1}{2i}(G(\omega+i\epsilon, l)-G(\omega-i\epsilon, l))= \pi \rho_{\omega}(\omega, l)\ .
\end{equation}
Substituting \eqref{eqrho1} into \eqref{eqrho2}, we find,
\begin{equation}
    \text{Im}G_R(\omega, l)=- \text{Im} G_R^{BH} (\omega, l) \bigg\rvert_{\omega=\omega_k} \ .
\end{equation}
It is worth emphasizing that, while this equation appears simple, it encodes a very non-trivial relation. The pole structures of the two Green's functions $G_R$ and $G_R^{BH}$, are fundamentally different. The Green's function $G_R$ has poles on the real axis, which correspond to the normal modes of the system. On the other hand,  $G_R^{BH}$ has poles in the lower-half complex plane, corresponding to black hole quasi normal modes. This equation indicates that, when the real-valued poles become very dense, they can be approximated as forming a branch cut. The discontinuity across this branch cut contains information about the complex-valued quasi normal modes.

\subsection{In position space}
So far, we have discussed the momentum space Green's function. In this section, we turn our attention to the position space Green's function, focusing primarily on the Feynman propagator, which is given by \footnote{Here $u, v$ are the position space variables conjugate to $w=\frac{l-\omega}{2}$ and $p=\frac{l+\omega}{2}$.}:
\begin{align}
    C(u, v) &= \mathcal{N} \iint dw dp \, e^{ipv-i w u} G_F(w, p) \nonumber \\
    &= \mathcal{N} \iint dw dp \, e^{ipv-i w u} \sum_{w_k} \left(\frac{w^2}{w_k^2}  \right)^{\Delta}\frac{\text{Res}(G, w_k)}{w-w_k\pm i\epsilon}\ .
\end{align}
Here, the $\pm$ sign corresponds to positive and negative $w_k$, respectively. For simplicity, we will consider $u>0$ without loss of generality, which allows us to select only the positive $w_k$. After integrating over $w$, the expression simplifies to:
\begin{equation}
     C(u, v)=\mathcal{N} \sum_{w_k}\int dp \, e^{ipv-i w_k u}  \text{Res}(G, w_k)\ .
\end{equation}
As mentioned earlier, when the brick wall is very close to the horizon, the sum over $k$ can be approximated as an integral. In this approximation, the discrete $w_k$ becomes a continuous variable $w$. Thus, the expression for $C(u, v)$ becomes \footnote{Here we are absorbing various numerical factors within $\mathcal{N}$ in various steps.}:
\begin{align}
    C(u, v) &=\mathcal{N} \iint dp dw \, e^{ipv-i w u}  \text{Im} G_R^{BH} (w, l) \nonumber \\
    & = \mathcal{N} \iint dp dw \, e^{ipv-i w u}  (G_R^{BH} (w, l)-G_R^{BH} (w, l)^*)\ .
\end{align}
Here, the first term has poles in the lower half-plane, while the second term has poles in the upper half-plane. Consequently, only the first term contributes, and it reproduces the exact position-space Green's function, which corresponds to that of a thermal one with a temperature equal to the Hawking temperature of the black hole.\\

This result implies that, although we start from a pure state, the two-point correlator can be well approximated by a thermal correlator when the brick wall is close to the horizon. Therefore, a boundary observer cannot distinguish this pure state from a thermal state unless they wait for an exceptionally long time or measure higher-point correlators.

\section{Discussion}\label{sec:7}

In this study, we explored a minimally coupled scalar field in the bulk five-dimensional AdS-Schwarzschild spacetime, which is dual to the gauge-invariant composite scalar operators  Tr($F_{\mu \nu}F^{\mu \nu}$) or Tr($F_{\mu \nu}\tilde{F}^{\mu \nu}$) in the boundary four dimensional $\mathcal{N}=4$ SYM theory. The scalar field was quantized using a Dirichlet boundary condition, ensuring that the scalar field vanishes on the stretched horizon, effectively imposing a perfectly reflecting boundary condition instead of the more conventional ingoing boundary condition near the horizon. This quantization results in normal modes that are real-valued, in contrast to the complex-valued quasi-normal modes typically observed. These normal modes are labeled by two quantum numbers: the principal quantum number $n$ and the angular momentum quantum number $l$. A significant degeneracy exists in the spectrum; for each $l$, there can be $(2l+1)^2$ states with the same $\omega$. Since the radial equation takes the form of a Heun equation, no exact solution is available, necessitating the use of various approximate methods to determine the normal modes. We employed techniques from Liouville CFT, as previously utilized in \cite{Dodelson:2022yvn}, to calculate these normal modes. Although the spectrum shows linearity along the $n$ direction, a non-trivial dependence is observed along the $l$ direction. However, the corresponding single-particle spectral form factor (SFF) does not exhibit a clear ramp structure. This can be attributed to the absence of logarithmic dependence of the low-lying modes along the $l$-direction, a feature that was crucial in the BTZ black hole case for producing the linear ramp. Furthermore, we were unable to obtain the modes for sufficiently large quantum numbers and for small $r_H/l\geq1$. Improving numerical methods to identify these could be beneficial, not only for SFF analysis but also for extracting black hole quasi-normal modes. Insights from these modes could provide a deeper understanding of strongly coupled thermal $\mathcal{N}=4$ super Yang-Mills (SYM) theory.

We then applied the WKB method to solve the spectrum, which allowed us to easily compute the high-lying modes. Using this approach, we found that the corresponding single-particle spectral form factor (SFF) along $l$ direction exhibits a clear Dip-Ramp-Plateau (DRP) structure, with a linear ramp of slope one in the log-log plot (see Figure \ref{WKB_sff}). It is important to note that the spectrum is deterministic, rather than random as seen in various Gaussian ensembles while choosing the matrix elements. Despite this determinism, the slope of the ramp is one, similar to that observed in random matrix theory\footnote{For a comprehensive overview of deterministic sequences that exhibit a linear ramp in the SFF, see \cite{Das:2023yfj}. Among these, the logarithmic sequence is the simplest example, demonstrating a linear ramp with a slope of one.}. However, it is important to note that our numerical methods currently encounter difficulties as we approach very close to the horizon, where the numerics break down for extremely small values of the cut-off $\epsilon_0$. It is important to note that while the SFF exhibits a distinct DRP structure, the LSD does not follow the conventional Wigner-Dyson distribution. However, by generalizing this simple boundary condition, as demonstrated in \cite{Das:2023ulz}, one can achieve a Wigner-Dyson LSD accompanied by a linear ramp.

We have attempted to solve the Heun equation perturbatively to determine the normal modes and the corresponding SFF. So far, we have only used the zeroth order equation to compute the normal modes. Although it is, in principle, possible to find higher-order corrections to the wave functions and normal modes, it is quite challenging in practice. Despite this, we have been able to show a linear ramp of slope \textit{one} the SFF along $l$-direction.

Since chaos is closely related to thermality, we expect observing thermal behavior emerging from the system. This behavior is demonstrated by computing the Green's function and showing that the Green's function of this pure state can be approximated by the thermal retarded Green's function of $\mathcal{N}=4$ SYM at a temperature equal to the black hole's Hawking temperature. This approximation holds when the brick wall is placed very close to the horizon, where the poles of the Green's function become so densely packed that they can be approximated as a branch cut. The discontinuity around this branch cut captures the thermal correlator's information. However, over long time scales, the pole structure becomes apparent instead of the branch cut. In \cite{Burman:2023kko}, it is suggested that this transition occurs around the Page time, approximately $O(N)$, and it would be valuable to examine this timescale by computing the gap in the spectrum. For a more detailed discussion, see \cite{Suman}.

The main result of this article is the clear Dip-Ramp-Plateau (DRP) structure with a linear ramp of slope one in the single-particle SFF constructed from the normal mode spectrum along the $l$ direction for $1\leq r_H/l \ll \infty $. This behavior arises from the non-trivial dependence of the spectrum on $l$ quantum number when the brick wall is placed very close to the horizon. In this limit, the retarded correlator can be effectively approximated by a thermal correlator.

There are several promising directions for future work. One avenue is to compute the normal modes for charged and rotating black holes. Another direction is to generalize this framework to fermionic or gauge fields. A particular concern is the absence of the ramp for $r_H/l\gg 1$, which is due to the non-logarithmic dependence of the low-lying normal modes along $l$. At this limit, the approximation in equation (5.13) of \cite{Krishnan:2023jqn} breaks down for the low-lying modes, although it remains valid for the high-lying modes, which means a logarithmic dependence. Therefore, it is possible that the high-lying modes will exhibit the ramp structure in the SFF, suggesting that the part of the spectrum responsible for the ramp may change as we vary $r_H/l$. Further exploration of this aspect would be very interesting.

Another issue concerns the $\Delta$-dependence of the modes depicted in Figure \ref{delta_dep}. We do not yet have an explanation for why this behavior is reversed compared to the BTZ case. If the large redshift (with a single zero of the blackening factor at the horizon) is responsible for this dependence of the modes, there should be no reason for this reversed $\Delta$-dependence (note that the $\Delta$-dependence of the WKB spectrum is similar to that observed in BTZ). Our hunch is that this behavior may be attributed to the parameter $a$ appearing in equation \eqref{MFa}, which becomes imaginary for the low-lying modes when $r_H>L$, but remains real in the limit $r_H\ll L$. The Liouville CFT techniques are well-suited for the small $t$ regime, i.e., $r_H \ll L$, which lies below the Hawking-Page transition point. However, in this analysis, we are working in a regime where $r_H>L$, although not as extreme as the black brane case where $t=1/2$. It is crucial to gain a better understanding of this behavior.

Additionally, an interesting open question arises from the observation in \cite{Hartnoll:2005ju} that the retarded Green's function of thermal SYM exhibits branch cuts in the lower half-plane in the weak coupling limit, where the bulk picture is highly quantum. In contrast, in the strong coupling limit, which corresponds to a classical black hole geometry, the Green's function contains complex-valued poles that are identified as quasi-normal modes. In our study, we observed that placing a brick wall in the geometry leads to poles that manifest as a branch cut when the wall is very close to the horizon. It is important to emphasize that these poles lie on the real line since there is no dissipation in the system. If, instead of a Dirichlet boundary condition, we introduce some loss by imposing a Neumann boundary condition (a \textit{not-so-hard} brick wall), it is conceivable that the poles could shift away from the real line, still manifesting as a branch cut when the brick wall is close to the horizon. We plan to address this issue in the future.

In a recent work \cite{Das:2024mlx}, the author has conjectured a correspondence between the excited states of the free scalar field in this brick-wall geometry to that of the vacuum or excited states of a CFT under modular quantization\cite{Tada:2019rls} (quantization for the modular Hamiltonian). Additionally,  in a separate study \cite{Das:2024vqe}, it was observed that a particular class of Floquet CFT dynamics corresponds to the extended modular Hamiltonian of a subsystem between two fixed points of the dynamics and the near horizon Virasoro algebra resembles the Virasoro algebra under Modular quantization. These observations prompt an intriguing future direction to check the validity of this conjecture (or a similar one) in higher dimensions (especially in the context of AdS$_5$/CFT$_4$). In particular, one might use the results of \cite{Das:2023xaw}, which generalizes the study of floquet CFTs in d>2, to check this in higher dimensions.

\section{Acknowledgements}

We express our sincere gratitude to Alexander Zhiboedov, Arnab Kundu, Bobby Ezhuthachan, Chethan Krishnan and  Matthew Dodelson for their valuable discussions and insightful comments.

\appendix

\section{Spectral form factor}\label{apnAA}

The Spectral Form Factor (SFF) is defined as follows:
\begin{equation}
    \mathcal{S}(t)=\left| \frac{\text{Tr} (e^{-(\beta+it)H}) }{\text{Tr} (e^{-\beta H}) } \right|^2 \, .
\end{equation}
If the energy spectrum is known, this can be rewritten as :
\begin{equation}
    \mathcal{S}(t)= \frac
    {\sum_{n} e^{-(\beta+it)E_n} \sum_{n'} e^{-(\beta-it)E_{n'}}} {(\sum_{n} e^{-\beta E_n})^2} \, .
\end{equation}

Instead of referencing generic facts on the SFF (see \cite{Cotler:2016fpe} for this), let us highlight some important aspects. First, when working with an infinite-dimensional Hilbert space and a known numerical spectrum, it is often impractical to sum over the entire spectrum. Instead, we can sum up to a certain cutoff, $n_{cut}$ , and then vary $n_{cut}$ to check whether $\mathcal{S}(t)$ remains stable.

Second, since $\vec{m}$ does not appear in the radial equation, the spectrum exhibits a large degeneracy. For each value of $l$, there are $(2l+1)^2$ number of modes with the same value. Consequently, the SFF becomes:
\begin{equation}
    \mathcal{S}(t)= \frac
    {\sum_{n,l_1} (2l_1+1)^2 e^{-(\beta+it)E_{n,l_1}} \sum_{n', l_2} (2l_2+1)^2 e^{-(\beta-it)E_{n',l_2}}} {(\sum_{n, l} (2l+1)^2 e^{-\beta E_{n, l}})^2} \, .
\end{equation}
Here, we will disregard these degeneracies and adopt the following definition for the SFF (symmetry resolved in some sense). The SFF along $n$-direction is:
\begin{equation}\label{sffn}
    \mathcal{S}(t) \Bigg|_{l, \vec{m}=\text{const.}}= \frac
    {\sum_{n} e^{-(\beta+it)E_n} \sum_{n'} e^{-(\beta-it)E_{n'}}} {(\sum_{n} e^{-\beta E_n})^2} \, .
\end{equation}
and SFF along $l$-direction becomes,
\begin{equation}\label{sffl}
    \mathcal{S}(t) \Bigg|_{ n, \vec{m}=\text{const.}}= \frac
    {\sum_{l_1} e^{-(\beta+it)E_{l_1}} \sum_{l_2} e^{-(\beta-it)E_{l_2}}} {(\sum_{l} e^{-\beta E_l})^2}
\end{equation}
In the main text, we focus on the two expressions above, which we have labeled as $g(t)$ in the various figures.

\section{WKB approximation method}\label{apnA}

In this section, we briefly review the WKB approximation method. For simplicity, we will restrict the discussions to one dimension. We start with the Schrödinger equation for the stationary states of a single particle moving in a time-independent potential,

\begin{equation}\label{sch}
    -\frac{\hbar^2}{2 m}\frac{d^2 \psi(x)}{d x^2}+ V(x) \psi(x)= E \psi(x)\, ,
\end{equation}
and consider $p(x)= \sqrt{2 m (E-V(x))}$.
If the potential is constant, the solutions of \eqref{sch}, for $E>V(x)$, are \footnote{For $E<V(x)$, the solutions would be $\psi(x) \sim B_{\pm} e^{\frac{\pm  p x}{\hbar} }$}
\begin{equation*}
    \psi(x) \sim A_{\pm} e^{\frac{\pm i p x}{\hbar} }\, .
\end{equation*}

However, if the potential is not constant, the solution is not as straightforward. However, we assume an ansatz for the solution as follows.

\begin{equation}\label{ansatz}
    \psi(x)= C(x)_{\pm} e^{\frac{\pm i \phi(x)}{\hbar} }\, .
\end{equation}

Substituting \eqref{ansatz} into \eqref{sch} and separating the real and imaginary parts, we obtain two differential equations:

\begin{eqnarray}\label{wkb}
    C''(x)- C(x) \phi'(x)^2+p(x)^2 C(x)=0\, , \\
    (C'(x)^2 \phi'(x))'=0 \, .
\end{eqnarray}

The first of the two equations above cannot be solved in general. However, we assume that the amplitude $C(x)$ is slowly varying, i.e. $C''(x)$ can be neglected. Under this assumption, the equation can be solved, and in the `classically allowed region' $(E>V(x))$, the wave function takes the following form:

\begin{equation}\label{CA}
    \psi(x)= \frac{A_{1}}{\sqrt{p(x)}} e^{\frac{i}{\hbar}\int_x^{x^*} p(x') dx'}+\frac{B_{1}}{\sqrt{p(x)}} e^{-\frac{i}{\hbar}\int_x^{x^*} p(x') dx'}\, ,
\end{equation}

whereas, in the `classically forbidden region' $(E< V(x))$, it is given by:

\begin{equation}\label{CF}
    \psi(x)= \frac{A_2}{\sqrt{|p(x)|}} e^{-\frac{1}{\hbar}\int^x_{x^*} |p(x')| dx'}+\frac{B_2}{\sqrt{|p(x)|}} e^{\frac{1}{\hbar}\int^x_{x^*} |p(x')| dx'}\, ,
\end{equation}
where, $A_{1,2},B_{1,2}$ are two sets of arbitrary constants that are yet to be fixed.

As we can see from above, there is a significant problem at the `classical turning points' where $E=V(x)$. At these points, both \eqref{CA} and \eqref{CF} diverge. Therefore, these points must be treated separately. The solutions at the turning points allow us to connect the constants of the solutions in \eqref{CA} and \eqref{CF}, as the turning point acts as a bridge between the two regions. We assume that the potential varies slowly near the turning point such that

\begin{equation}\label{pot}
    V(x)= E+ V'(x^*)(x-x^*) \, ,
\end{equation}
where, $x^*$ is the classical turning point which is derived by solving $V(x)=E$.\\
If we substitute \eqref{pot} in \eqref{sch}, the Schrödinger equation becomes,

\begin{equation}
    \frac{d^2 \psi(x)}{dx^2}= \zeta^3 (x-x^*) \psi(x)\, ,
\end{equation}
with $\zeta= \left(\frac{2 m}{\hbar^2}V'(x^*)\right)^{1/3}$. After a change of variable, $z \equiv \zeta (x-x^*)$, the above equation takes the form of the Airy's equation:
\begin{equation*}
    \frac{d^2 \psi(z)}{dz^2}=z \psi(z)\, .
\end{equation*}

Therefore, the solutions are given as a sum of the Airy functions:

\begin{equation}\label{airy}
    \psi(z)= \alpha Ai(z) +\beta Bi(z) \, .
\end{equation}

\begin{figure}
\centering
\begin{tikzpicture}
    \draw[-latex] (-0.5,0) -- (10,0) node[right] {$x$};
    \draw[-latex] (0,-0.5) -- (0,5) node[above] {$V(x)$};
    
    \draw[thick] plot[domain=0.5:10, smooth] (\x,{4*(1-exp(-0.5*\x))});
    \draw[dashed] plot[domain=0.1:0.5] (\x,{4*(1-exp(-0.5*\x))});
    
    \draw[dashed] (0,3) -- (10,3) node[right] {$V(x)=E$} ;
    
    \draw[dashed] (2.8,0) -- (2.8,5);
    \node[below] at (2.8,0) {$x*$};
    
    \node at (1.4,1) {I};
    \node at (6.4,1) {II};
    
    \draw (-0.5,-0.5) rectangle (10,5);
    
\end{tikzpicture}
\caption{Schematic diagram of a potential, illustrating the `classically allowed' and `classically forbidden' regions for a particle with energy E. The point $x*$ denotes the turning point.}
\label{wkbpot}
\end{figure}
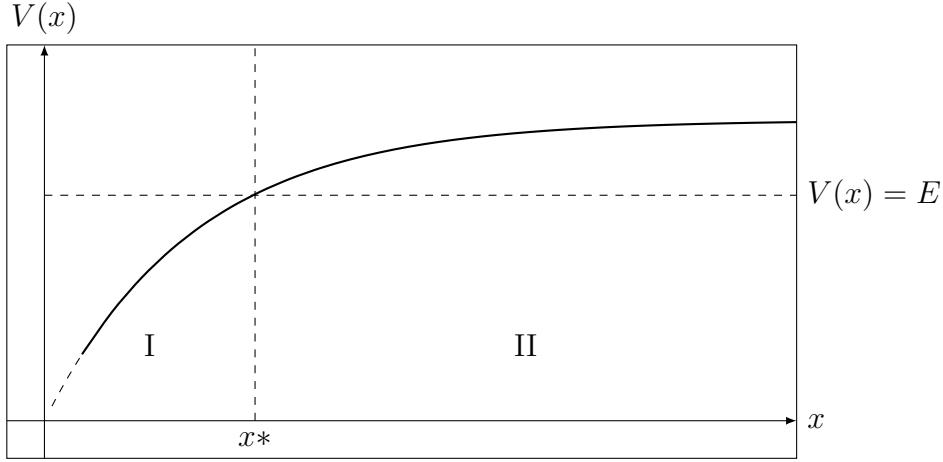

Let us consider a scenario (Figure [\ref{wkbpot}]) to illustrate how this method is used in practice. Assume that the potential is such that the region where
$-\infty<x<x^*$ is classically forbidden (region II), and the region where  $\infty>x>x*$ is classically allowed (region I). In this case, the wave function in region I is given by \eqref{CA}, and in region II, it is given by \eqref{CF}. Since the asymptotic forms of both $Ai(z)$ and $Bi(z)$ are known, by matching \eqref{airy} with \eqref{CA} and \eqref{CF} in the two asymptotic regions $z \gg 0$ and $z \ll 0$ we can derive the WKB connection formulae relating the constants of the solutions $\{A_1,B_1\}$, $\{A_2,B_2\}$ and $\{\alpha,\beta\}$. 

In particular, we can consider an example of the type Figure [\ref{WKB_pot1}] with one turning point. Then, using the normalizability condition for solution in $z \gg 0$ region, which implies $B_2=0$, one can show that the connection formulae are given by:

$$A_1=-i e^{i\frac{\pi}{4}} A_2, B_1= i e^{-i\frac{\pi}{4}} A_2 \, ,$$ and correspondingly the solutions in the two regions are given by:
\[ \psi(x)=
\begin{cases}
    \frac{2 A_2}{\sqrt{p(x)}}\cos\left(\frac{1}{\hbar}\int_x^{x^*} p(x') dx'-\frac{\pi}{4} \right), \quad \ \text{for region-I}\\ 
    \frac{A_2}{\sqrt{|p(x)|}}\exp\left(-\frac{1}{\hbar}\int^x_{x^*} |p(x')| dx' \right), \quad \ \text{for region-II} \, .
\end{cases}
\]

This type of potential has a cut-off at  $\psi(x_0)=0$, for some $x_0<x^*$, then from the solution for region-II, one finds that $\cos\left(\frac{1}{\hbar}\int_{x_0}^{x^*} p(x') dx'-\frac{\pi}{4} \right)$ must be zero for $\psi(x_0)=0$, in other words, the condition 
$$\frac{1}{\hbar}\int_{x_0}^{x^*} p(x') dx'=\frac{3 \pi}{4} +n \pi,\ \quad \forall n\in \mathbb{Z}$$
must be satisfied.\\

\noindent \textbf{Validity of the WKB approximation:}\\

We will end this section by discussing the validity of the WKB approximation method. To do that, we put $\psi(x)= e^{i \frac{\phi(x)}{\hbar}}$ in \eqref{sch} and we get 

\begin{equation}
    i \hbar \phi''(x)- (\psi'(x))^2+ p(x)^2=0
\end{equation}
Therefore, in the quasi-classical limit one would expect

\begin{eqnarray*}
    |\hbar \phi''(x)|<< (\psi'(x))^2 \\
   or, \left|\frac{d}{dx}\left(\frac{\hbar}{\phi'(x)}\right) \right|<<1 
\end{eqnarray*}

Since, from \eqref{wkb} we know that $\phi'(x)=p(x)$ we can write the condition as

\begin{equation}
    \left|\frac{d}{dx}\left(\frac{\hbar}{p(x)}\right) \right|<<1 
\end{equation}

From the above equation, we can see that, at the classical turning point, the condition does not hold since $p(x)=0$, as we mentioned earlier also.

\section{Connection formulae}\label{apnB}
Following the connection formulas given in \cite{Bonelli:2022ten} and \cite{Bhatta_2023}, the incoming and outgoing modes near the horizon can be written in terms of the normalizable and non-normalizable modes near the boundary as

\be
\begin{gathered}\label{b1}
    \chi^{(\tilde{t})}_{+}= \sum_{\theta'=\pm} \left(M_{++}(a_{\tilde{t}}, a; a_0) M_{-\theta'}(a, a_1; a_{\infty}) \tilde{t}^a e^{-\frac{1}{2}\partial_a F}+ M_{+-}(a_{\tilde{t}}, a; a_0) M_{+\theta'}(a, a_1; a_{\infty}) \tilde{t}^{-a} e^{\frac{1}{2}\partial_a F}\right) \\\tilde{t}^{\frac{1}{2}-a_0+a_{\tilde{t}}} (1-\tilde{t})^{a_t-a_1} e^{\frac{1}{2}(\partial_{a_{\tilde{t}}}+ \theta' \partial_{a_1})F} \chi^{(1)}_{\theta'}
\end{gathered}
\ee
and

\be
\begin{gathered}\label{b2}
    \chi^{(\tilde{t})}_{-}= \sum_{\theta'=\pm} \left(M_{-+}(a_{\tilde{t}}, a; a_0) M_{-\theta'}(a, a_1; a_{\infty}) \tilde{t}^a e^{-\frac{1}{2}\partial_a F}+ M_{+-}(a_{\tilde{t}}, a; a_0) M_{+\theta'}(a, a_1; a_{\infty}) \tilde{t}^{-a} e^{\frac{1}{2}\partial_a F}\right) \\\tilde{t}^{\frac{1}{2}-a_0+a_{\tilde{t}}} (1-\tilde{t})^{a_{\tilde{t}}-a_1} e^{\frac{1}{2}(\partial_{a_{\tilde{t}}}+ \theta' \partial_{a_1})F} \chi^{(1)}_{\theta'}
\end{gathered}
\ee
where,
\begin{eqnarray}\label{MFa}
   & M_{\theta \theta'}(\alpha, \beta; \gamma) =\frac{\Gamma(-2 \theta'\beta)\Gamma(1+2 \theta\alpha)}{\Gamma(\frac{1}{2}+\theta\alpha-\theta'\beta+ \gamma) \Gamma(\frac{1}{2}+\theta\alpha-\theta'\beta-\gamma)}\\
   & F(\tilde{t}) = \frac{(\frac{1}{4}-a^2-a^{2}_1+a^{2}_{\infty})(\frac{1}{4}-a^2-a^{2}_{\tilde{t}}+a^{2}_{0})\tilde{t}}{\frac{1}{2}- 2a^2}\\\label{MFal}
   & a^2 = \left(-\frac{1}{4}-u+a^{2}_{\tilde{t}}+a^{2}_0)\right)\left(1-\frac{(-1+2 a^{2}_0+2 a^{2}_1-2 a^{2}_{\infty}+2 a^{2}_{\tilde{t}}-2 u)(-1+ 4 a^{2}_{\tilde{t}}-2 u)}{2 (-1+4 a^2_{0}+4 a^2_{\tilde{t}}-4 u) (-1+2a_0^2+2a_{\tilde{t}}^2-2u)}\tilde{t}\right)^2
\end{eqnarray}
In the above, we used the expansion of the conformal block 
F in the parameter t \cite{Bonelli:2022ten}, retaining only the leading-order term.
The values of the parameters depend on the specific space-time that we are considering. For the AdS-Schwarzschild black hole, the parameters are \cite{Bhatta_2023}

\begin{eqnarray}\label{bhprmtr}
    \tilde{t} &=& \frac{r^2_{h}}{2 r^2_{h}+1} \\
    a_{\tilde{t}} &=& \frac{i \omega}{2} \frac{r_h}{2 r^2_{h}+1} \\
    a_1 &=& \frac{\Delta-2}{2}\\
    a_0 &=& 0\\
    a_{\infty} &=& \frac{\omega}{2} \frac{\sqrt{r^2_{h}+1}}{2 r^2_{h}+1} \\ \label{bhprmtr1}
    u &=& - \frac{l(l+2)+2 (2 r^2_{h}+1)+ r^2_{h} \Delta (\Delta-4)}{4 (r^2_{h}+1)} + \frac{r^2_{h}}{4 (1+ r^2_{h})} \frac{\omega^2}{2 r^2_{h}+1}
\end{eqnarray}

Substituting \eqref{b1} and \eqref{b2} in \eqref{lcft2}, we extract the parameters $a_{11}, a_{22}, b_{11}$ and $b_{22}$ to be:

\begin{eqnarray}\label{paramtrs}
 a_{11}=[M_{++}(a_{\tilde{t}}, a; a_0) M_{-+}(a, a_1; a_{\infty}) \tilde{t}^a e^{-\frac{1}{2}\partial_a F}+M_{+-}(a_{\tilde{t}}, a; a_0) M_{++}(a, a_1; a_{\infty}) \tilde{t}^{-a} e^{\frac{1}{2}\partial_a F}] \nonumber\\
    \tilde{t}^{\frac{1}{2}-a_0+a_{\tilde{t}}} (1-\tilde{t})^{a_t-a_1} e^{-\frac{1}{2}(\partial_{a_{\tilde{t}}}+\partial_{a_1})F}\nonumber\\
a_{22}=[M_{++}(a_{\tilde{t}}, a; a_0) M_{--}(a, a_1; a_{\infty}) \tilde{t}^a e^{-\frac{1}{2}\partial_a F}+M_{+-}(a_{\tilde{t}}, a; a_0) M_{+-}(a, a_1; a_{\infty}) \tilde{t}^{-a} e^{\frac{1}{2}\partial_a F}]\nonumber \\
    t^{\frac{1}{2}-a_0+a_{\tilde{t}}} (1-\tilde{t})^{a_{\tilde{t}}-a_1} e^{-\frac{1}{2}(\partial_{a_{\tilde{t}}}-\partial_{a_1})F}\nonumber\\
b_{11}=[M_{-+}(a_{\tilde{t}}, a; a_0) M_{-+}(a, a_1; a_{\infty}) \tilde{t}^a e^{-\frac{1}{2}\partial_a F}+M_{--}(a_{\tilde{t}}, a; a_0) M_{++}(a, a_1; a_{\infty}) \tilde{t}^{-a} e^{\frac{1}{2}\partial_a F}]\nonumber \\
    \tilde{t}^{\frac{1}{2}-a_0-a_t} (1-\tilde{t})^{a_{\tilde{t}}-a_1} e^{-\frac{1}{2}(\partial_{a_{\tilde{t}}}+\partial_{a_1})F}\nonumber\\
b_{22}=[M_{-+}(a_{\tilde{t}}, a; a_0) M_{--}(a, a_1; a_{\infty}) \tilde{t}^a e^{-\frac{1}{2}\partial_a F}+M_{--}(a_{\tilde{t}}, a; a_0) M_{+-}(a, a_1; a_{\infty}) \tilde{t}^{-a} e^{\frac{1}{2}\partial_a F}] \nonumber\\
    \tilde{t}^{\frac{1}{2}-a_0-a_{\tilde{t}}} (1-\tilde{t})^{a_{\tilde{t}}-a_1} e^{-\frac{1}{2}(\partial_{a_{\tilde{t}}}-\partial_{a_1})F}\nonumber\\
\end{eqnarray}

 Substituting the parameters for AdS-Schwarzchild black hole \eqref{MFa}-\eqref{bhprmtr1} in \eqref{paramtrs}, one can check that the ratio of the parameters $\frac{b_{22}}{a_{22}}$ can be written as:
\begin{equation}\label{absolt}
    \frac{b_{22}}{a_{22}}= (\tilde{t})^{-i Q}\frac{\Lambda}{\Lambda^*}
\end{equation}
 where, $\tilde{t}=\frac{r_h^2}{2 r_h^2+1}$, $Q=\frac{r_h \omega }{2 r_h^2+1}$, $\Lambda=\Gamma(1-i Q)\big(e^{P_1} R_1 \Gamma \left(\frac{1}{2}-a-i Q\right)^2+e^{P_2} R_2\Gamma \left(\frac{1}{2}+a-i Q\right)^2 \big)$ and
$P_1=\frac{a}{\left(1-4 a^2\right)^2}\left(-4 a^2+\frac{r_h^2 \omega ^2}{\left(2 r_h^2+1\right)^2}+1\right) \left(-4 a^2-(\Delta -2)^2+\frac{\left(r_h^2+1\right) \omega ^2}{\left(2 r_h^2+1\right)^2}+1\right)$,\\ 

$R_1=\Gamma (2 a) \Gamma (2 a+1) \Gamma \left(\frac{1}{2} \left(-2 a+\Delta -\frac{\sqrt{r_h^2+1} \omega }{2 r_h^2+1}-1\right)\right) \Gamma \left(\frac{1}{2} \left(-2 a+\Delta +\frac{\omega  \sqrt{r_h^2+1}}{2 r_h^2+1}-1\right)\right)$,\\

$P_2=\frac{a \left(8 a^2 \left(r_h^2+1\right)+\Delta^2-4 \Delta +2 \left(\Delta ^2-4 \Delta +2\right) r_h^2-\omega ^2+2\right)}{\left(4 a^2-1\right) \left(2 \text{rh}^2+1\right)}$,\\

$R_2=\Gamma (-2 a) \Gamma (1-2 a) \Gamma \left(\frac{1}{2} \left(2 a+\Delta -\frac{\sqrt{r_h^2+1} \omega }{2 r_h^2+1}-1\right)\right) \Gamma \left(\frac{1}{2} \left(2 a+\Delta +\frac{\omega  \sqrt{r_h^2+1}}{2 r_h^2+1}-1\right)\right)$.
Therefore, from \eqref{absolt}, it is clear that $|\frac{b_{22}}{a_{22}}|=1$.


\bibliography{bibliography1.bib}
\bibliographystyle{JHEP.bst}





\end{document}